\documentclass[12pt]{article}
\usepackage[fontsize=12pt]{fontsize}
\usepackage[left=1.25in,right=1.25in,top=1.25in,bottom=1.25in]{geometry}
\usepackage{setspace}
\usepackage[title]{appendix}
\usepackage{enumitem}
\usepackage{tikz}
\usepackage{multirow}
\usepackage{subcaption}
\usepackage{ragged2e}
\usepackage{lmodern}

\usepackage{amsmath,amssymb,amsfonts,amsthm,bbm,mathrsfs,tcolorbox,tikz}
\usepackage{natbib}
\usepackage{cancel,sectsty,color,enumitem,array,hyperref,graphicx,subcaption}
\usepackage[justification=centering]{caption}
\hypersetup{pdfborder = {0 0 0},colorlinks=true,linkcolor=blue,urlcolor=blue,citecolor=blue}
\usepackage[normalem]{ulem}
\usepackage{booktabs}
\usepackage{rotating}
\usepackage{tabularx}

\usepackage{titlesec}

\titlespacing*\section{0pt}{4pt plus 4pt minus 2pt}{0pt plus 2pt minus 2pt}
\titlespacing*\subsection{0pt}{4pt plus 4pt minus 2pt}{0pt plus 2pt minus 2pt}
\titlespacing*\subsubsection{0pt}{4pt plus 4pt minus 2pt}{0pt plus 2pt minus 2pt}

\newtheoremstyle{exampstyle}
  {} 
  {} 
  {} 
  {} 
  {\bfseries\color{blue}} 
  {.} 
  {.5em} 
  {} 
\theoremstyle{exampstyle}\newtheorem{rem}{Remark}
\theoremstyle{exampstyle}
\theoremstyle{exampstyle}\newtheorem*{thm*}{Theorem}
\theoremstyle{exampstyle}\newtheorem{defn}{Definition}     
\theoremstyle{exampstyle}     
\theoremstyle{exampstyle}\newtheorem{lemma}{Lemma}  
\theoremstyle{exampstyle}\newtheorem{coro}{Corollary}        
\theoremstyle{exampstyle}\newtheorem{prop}{Proposition}
\theoremstyle{exampstyle}  
\theoremstyle{exampstyle}

\theoremstyle{exampstyle}  
\theoremstyle{exampstyle}
\theoremstyle{exampstyle}
\theoremstyle{exampstyle}\newtheorem{ax}{Axiom}

\newtheorem{myexp}{Example}

\theoremstyle{exampstyle}

\theoremstyle{definition}

\allowdisplaybreaks[1]

\usepackage{pifont}

\newcommand{\xy}{\{x,y\}}

\usepackage{array}
\newcolumntype{M}[1]{>{\centering\arraybackslash}m{#1}}
\newcolumntype{N}{@{}m{0pt}@{}}

\title{Random Collection\thanks{I am indebted to Yusufcan Masatlioglu, Emel Filiz-Ozbay, and Erkut Y. Ozbay for their continuous guidance, encouragement, and support. I want to thank Marina Agranov, Victor H. Aguiar, Yunus C. Aybas, Christopher P. Chambers, Kfir Eliaz, Andrew Ellis, Onur Kesten, Matthew Kovach, Marco Mariotti, Kirby Nielsen, Quang Pham, Daniel Reck, Gerelt Tserenjigmid, Daniel Vincent, Kemal Yildiz, and seminar audiences at the University of Maryland, Caltech Summer Program, and BRIC XI for helpful discussions and comments. I am also grateful to the editor and two anonymous referees for their constructive comments and valuable suggestions, which have greatly improved the paper.}
}

\author{Tri Phu Vu\thanks{\protect\linespread{1}\protect\selectfont  Department of Economics, 3114 Tydings Hall, 7343 Preinkert Dr., College Park, MD 20742. Email: \texttt{tvuphu@umd.edu.}}\\ University of Maryland}

\vspace{2cm}
\date{Nov 3, 2026}

\begin{document}
\maketitle
{\fontsize{12}{14}\selectfont
\begin{abstract}
This paper studies choice situations in which a decision maker can choose multiple alternatives. Given a menu of available options, the decision maker selects a subset of the menu with certain probabilities. We employ an axiomatic approach to characterize various parametric models in the literature. We identify two main behavioral postulates that organize the models into overlapping classes based on their shared behavioral properties. Our results elucidate the implications of the functional form assumptions and shed light on the distinctions between models. The behavioral postulates offer simple tools for testing and falsifying the choice procedures used by the decision maker and reveal a close connection between models that are seemingly unrelated.

\end{abstract}
}

\noindent Keywords: Stochastic choice correspondence, choice bundles, choice correspondence, consideration set.

\noindent  JEL classification: D01, D11, D91.


\newpage
\section{Introduction}
\newcommand{\rev}[1]{{\color{red}#1}}

Decision makers frequently encounter choices with many options. The standard approach to modeling choice behavior assumes that decision makers select a single alternative each time they face a choice problem. In many environments, however, it is natural for decision makers to choose a collection of options rather than just one. Examples are abundant. Platforms often select a set of products to recommend to their customers. Supermarkets regularly choose an assortment of products to offer. Coaches in soccer or basketball typically pick a group of players for the starting lineup. Investors select a portfolio of financial assets to invest in. Investigating the behavior of decision makers in such situations is crucial, as it enables more accurate modeling of choices that naturally involve selecting multiple options.

This paper studies choice problems where decision makers can choose a collection of alternatives, allowing for variation in the selected collection. Such problems arise across a wide range of fields, including marketing, industrial organization, sports analytics, finance, and political science. Formally, given a set $S$ of feasible options, decision makers can choose an arbitrary subset $T$ of $S$ with a probability $\mu(T,S)$, where $\mu$ represents a probability measure over the power set of $S$. Various examples and interpretations of $\mu$ exist; we present several below.

\begin{itemize}
    \item \textbf{Product recommendation:} A platform uses an algorithm based on individual characteristics, purchase history, popularity of the products, etc., to recommend some products to customers from the set of products it offers. In online environments, Amazon's ``Amazon's Choice," Spotify's ``Editor's Picks," and Disney+'s ``Fan Favorites" are typical examples. In physical settings, \cite{kawaguchi2021designing} illustrate that some vending machines in Japan can utilize facial recognition technology to analyze a customer's age and gender and provide drink recommendations based on these demographic factors. Because the underlying data varies across customers and over time, the platform (or the vending machine) recommends each collection $T$ with probability $\mu(T,S)$.

    \item \textbf{Assortment choice:} A supermarket selects a collection of items to offer to its customers from the set of all products it can provide. Due to limited shelf space, variation in items' perishability, and fluctuating demand, the collection of items the supermarket selects varies over time. Consequently, the supermarket offers a set $T$ of products with probability $\mu(T,S)$. Such random assortments can arise from the supermarket optimizing its expected revenue \citep{ma2023assortment}.

    \item \textbf{Approval behavior:} A decision maker selects a collection of all options she approves. Here, $\mu(T,S)$ denotes the probability that the decision maker approves exactly the options in $T$ from menu $S$. One popular example is approval voting in which a voter reports a set of approved candidates without ranking them \citep{brams1978approval,fishburn1988experiment}. These approval ballots are used widely in political elections, sports awards, corporate governance, and association elections. Another example is the online approval behavior studied by \citet{manzini2024model}, where users select multiple items for later consideration, including adding products to a shopping cart or videos to a playlist.
\end{itemize}

Depending on the context, $\mu(T,S)$ may represent the choices of a single decision maker (intrapersonal interpretation) or those of different decision makers in a population (interpersonal interpretation). For instance, in approval voting, the analyst may observe ballots from multiple voters, and $\mu(T,S)$ may indicate the proportion of voters approving a set $T$ of candidates from the available candidates.

Various functional forms of $\mu$ have been proposed in the literature, each grounded in distinct underlying narratives and offering different interpretations. Rather than introducing a new model, this paper undertakes a systematic examination of well-known parametric models of $\mu$ listed in Table \ref{tab: models}.\footnote{Several models in Table \ref{tab: models} were originally proposed to study the formation of consideration sets, which is another interpretation of $\mu$ (see, for instance, \cite{Manzini-Mariotti_2014_ECMA} and \cite{cattaneo2020random}, among others). We will thoroughly illustrate in Section \ref{sec:models_and_characterization} that these parametric functional forms can also be applied to investigate choice situations that naturally involve selecting multiple options, such as supermarket assortment problems or product recommendations.} Despite their widespread use and numerous applications, little is known about their behavioral implications. Additionally, it is unclear how to distinguish these models from each other, as they often have complicated functional forms and are motivated by distinct stories. This paper provides behavioral characterizations of several formulations of $\mu$ in Table \ref{tab: models} and highlights a close connection between well-known but seemingly unrelated models. Our results offer a better understanding of the implications of functional form assumptions, allowing the analyst to differentiate between alternative models of $\mu$ and to identify the most suitable model for their specific application.

\begin{table}[ht!] 
\centering
\footnotesize
  \begin{tabular}{p{4.5cm}|p{4.3cm}|p{5.6cm}}
    \toprule
    \textbf{Models} & \textbf{Papers} &\textbf{Factors impacting $\mu$} \\
    \hline
    \textbf{List A} & & \\ 
    $\quad$ Logit & \cite{Brady-Rehbeck_2016_ECMA} & Importance (or salience) of sets of options \\
   $\quad$  Nested stochastic collection & \cite{kovach2022behavioral} & Nests (a partition of the grand set) \\
    $\quad$ Nested logit & \cite{mcfadden1978modelling} &Nests and the DM's preference \\
    \hline
   \textbf{List B} & & \\
    $\quad$ Independent Choice & \cite{Manzini-Mariotti_2014_ECMA} & Importance (or salience) of individual options \\
    \hline
   \textbf{List C} & & \\
    $\quad$ Random categorization & \cite{Aguiar_2017_EL} & Salience of categories of options \\
    $\quad$ Elimination by aspects &  \cite{Tversky:1972} & Salience of attributes of options \\
    $\quad$ Attribute rule & \cite{Gul_Natenzon_Pesendorfer_2014_ECMA} & Salience of attributes of options \\
    $\quad$ Random reference model & \cite{kibris2024random} & Salience of individual options \\
    \bottomrule 
  \end{tabular}
  \caption{Models of $\mu$ and their underlying motivations}
        \label{tab: models}
 \caption*{\footnotesize \justifying \textit{Notes.} Models are divided into lists A, B, and C based on their behavioral characterizations.}        
\end{table}
\normalsize

The paper's main contribution is to provide a behavioral taxonomy of stochastic collection-choice models that helps organize them into overlapping classes based on their shared behavioral properties. We identify two common behavioral postulates and their variants: Independence of Irrelevant Sets (IIS) and Relative Additivity. IIS states that the relative choice probability of two collections is independent of the presence or absence of other sets:
\[
\frac{\mu(T,S)}{\mu(T',S)}=\frac{\mu(T,S')}{\mu(T',S')} \quad \text{ for all non-empty } T,T' \subseteq S\cap S'.
\]
Meanwhile, roughly speaking, Relative Additivity states that adding a new option to the choice set has a constant effect on the choice probabilities of a given collection. That is, for a fixed $S\ni x$,
\[
 \frac{\mu(T,S\setminus \{x\})}{\mu(T,S)+\mu(T\cup \{x\}, S)} \quad \text{is independent of $T$ for all non-empty } T\subseteq S\setminus \{x\}.
\]
To understand Relative Additivity, suppose $S\setminus \{x\}$ is the initial choice set and $T$ is the chosen collection. Adding $x$ to the choice set can affect $T$ in two different ways. On the one hand, $x$ may have only an indirect effect, in which case it does not alter the chosen collection but only affects its selection probability. In that situation, $T$ is selected with probability $\mu(T,S)$. On the other hand, $x$ may have a direct impact and be added to the chosen collection, in which case $T\cup \{x\}$ is selected with probability $\mu(T\cup \{x\},S)$. Thus, the sum $\mu(T,S)+\mu(T\cup \{x\}, S)$ can be interpreted as the total choice probability associated with $T$ under the expanded choice set. The ratio $\frac{\mu(T,S\setminus \{x\})}{\mu(T,S)+\mu(T\cup \{x\}, S)}$ then captures the relative probability of choosing a collection before and after adding the new option. Relative Additivity states that this relative choice probability is independent of $T$. Section \ref{sec:models_and_characterization} discusses these two properties and their implications in greater detail.

The behavioral taxonomy and characterizations presented in the paper are useful in four different ways. First, many formulations of $\mu$, despite their cumbersome functional forms, have simple and intuitive behavioral foundations, with several characterized by a single axiom. While functional forms facilitate empirical applications, it is often challenging to tell whether two parametric formulations are equivalent, related, or distinct, especially when they have complicated forms. Identifying behavioral foundations allows us to decompose models into interpretable restrictions, organize them into overlapping classes with common implications, and perform such comparisons more easily. Specifically, in Section \ref{sec:models_and_characterization} we show that IIS and its variants appear in the characterizations of $\mu$ in Logit, Independent Choice, and Nested Stochastic Collection. Meanwhile, Relative Additivity and its variants appear in the characterizations of $\mu$ in Independent Choice, Random Categorization, Elimination by Aspects, Attribute Rule, and Random Reference Model. The Independent Choice model belongs to both classes, as it satisfies both main postulates. Comparing these behavioral restrictions also reveals that models motivated by distinct narratives can be behaviorally equivalent, while models with similar functional forms may intersect only in special cases.

Second, collection-choice models are widely used in various fields, including in marketing, industrial organization, health economics, transportation research, and social choice. Applications include approval voting
\citep{fishburn1988experiment}, airport choice
\citep{Basar:Bhat:2004}, physicians' hospital referrals \citep{gaynor2016free}, product recommendations
\citep{kawaguchi2021designing}, households' insurance deductible choices \citep{barseghyan2021discrete}, and
consideration of prescription drug insurance plans \citep{abaluck2021consumers}. The taxonomy presented in the paper thus helps analysts interpret and use models developed in other fields by clarifying the behavioral assumptions they bring into new applications and how their chosen specification relates to other models.

Third, the taxonomy guides the development of new, tractable collection-choice models that can accommodate additional choice irregularities such as context effects and violations of monotonicity. In Section \ref{sec:constructive}, we present one such modeling exercise by constructing other models that retain either IIS or Relative Additivity while relaxing the other, as well as an example that relaxes both. These examples clarify which additional choice patterns can arise when each postulate is relaxed, helping researchers decide which restrictions to retain and which to relax when extending a benchmark model to explain a particular choice irregularity.

Finally, the behavioral postulates identified offer a set of testable predictions that allow for empirical verification of these models. An outside analyst with observed data on $\mu$ can utilize these behavioral implications to identify which model classes are consistent with the observed data and potentially falsify others. Such data are available in several applications, including approval voting and product recommendation.\footnote{When $\mu$ is not directly observed, it can be elicited or estimated under suitable assumptions. See Section \ref{sec:observability} for a discussion of the observability of $\mu$.}

\begin{figure}[ht!]
    \centering
        \includegraphics[scale=0.20]{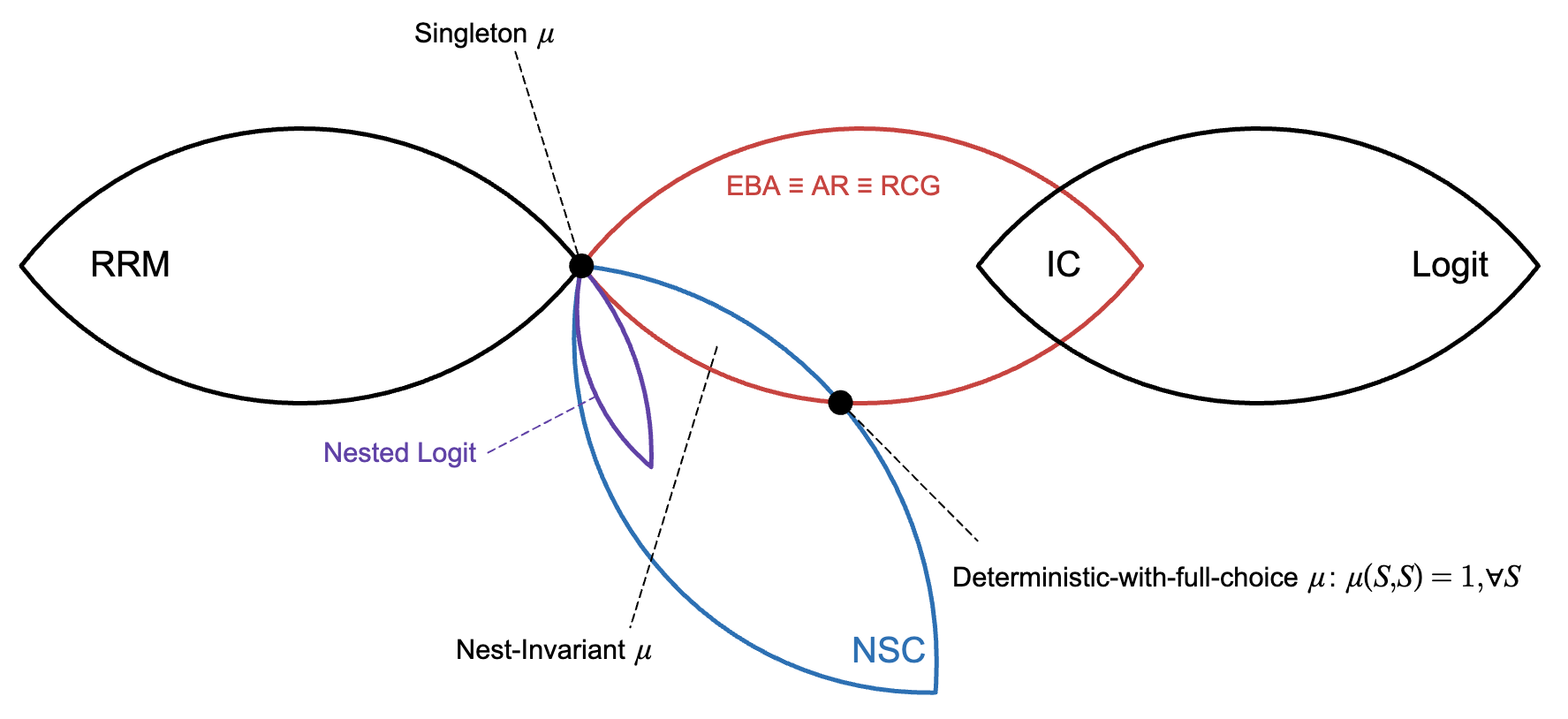}
        \caption{Relationship among various models of $\mu$}
        \label{fig: model relationship}
        \caption*{\footnotesize \justifying \textit{Notes.} Label legend: Logit \citep{Brady-Rehbeck_2016_ECMA}, IC---Independent Choice \citep{Manzini-Mariotti_2014_ECMA}, EBA---endogenous (static) Elimination By Aspects \citep{Tversky:1972}, AR---endogenous (first stage) Attribute Rule \citep{Gul_Natenzon_Pesendorfer_2014_ECMA}, RCG---Random Categorization \citep{Aguiar_2017_EL}, NSC---Nested Stochastic Collection \citep{kovach2022behavioral}, RRM---Random Reference Model \citep{masatlioglu2014canonical,kibris2024random}. The definitions of deterministic-with-full-choice $\mu$, singleton $\mu$, and nest-invariant $\mu$ are given in Section \ref{sec: relationship}. Singleton-$\mu$ is the intersection of RRM, NSC, and EBA/AR/RCG. Singleton-$\mu$ is related to the Luce model in the probabilistic choice literature. Nest-invariant $\mu$ is the intersection of EBA/AR/RCG and NSC. Nest-invariant $\mu$ is a subset of NSC satisfying the probabilistic version of the attention filter condition studied in \cite{Masatlioglu-Nakajima-Ozbay_2012_AER}. We assume that the chosen collection is non-empty in all models of $\mu$ in the figure.}        
\end{figure}

\medskip
\noindent \textbf{Relationship between models.} Figure \ref{fig: model relationship} summarizes the connections among various models of $\mu$. We identify the following key relationships. First, Random Reference Model is notably distinct from other frameworks: it intersects with some of them only in a special case of $\mu$ referred to as singleton $\mu$. This $\mu$, formally defined in Section \ref{sec: relationship}, is related to the Luce model in the probabilistic choice literature. Second, many models of $\mu$ intersect at the singleton $\mu$, but not at the deterministic-with-full-choice $\mu$, where the whole choice set is always selected with certainty. Third, the endogenous versions of the formulations of $\mu$ in Elimination by Aspects (static version, see Section \ref{sec:salience}), Attribute Rule, and Random Categorization are all equivalent. Finally, the formulation of $\mu$ in Independent Choice is equivalent to the intersection of Logit and Random Categorization, a result also obtained by \cite{kovach2023reference} in a slightly different setting. 

The paper proceeds as follows. Section \ref{sec:models_and_characterization} introduces various models of $\mu$ and provides their behavioral characterizations. Section \ref{sec: relationship} analyzes the relationships among these models. Section \ref{sec:constructive} uses the two behavioral postulates to introduce richer classes of collection-choice models. Section \ref{sec:discussion} examines the observability and other properties of $\mu$ and considers whether collection selection is followed by a second-stage choice. Section \ref{sec: Literature Review} reviews the related literature. Appendix \ref{appendix:additional_results_and_models} provides characterizations of additional models. All proofs are delegated to Appendix \ref{appendix: proofs}.


\section{Models and behavioral characterizations}\label{sec:models_and_characterization}

\subsection{Primitives} \label{sec: primitives}
\newcommand{\XO}{\mathcal{X}^o}
\newcommand{\SO}{\mathcal{S}^o}

Let $X$ be the finite set of all alternatives. We call $X$ the grand set and assume that there are at least three distinct alternatives in $X$. Typical elements of $X$ are denoted by $x,y,z$. Let $\mathcal{X}$ be the set of all non-empty subsets of $X$. Let $S\in \mathcal{X}$ denote a choice set. For notational simplicity, given a set $S$ and an element $x$, we write $S\cup x$ instead of $S\cup \{x\}$ and $S\setminus x$ instead of $S\setminus \{x\}$. Denote the sets of non-negative and positive real numbers by $\mathbb{R}_+$ and $\mathbb{R}_{++}$, respectively. 

Let $\mu$ denote a stochastic choice correspondence from $\mathcal{X} \times \mathcal{X}$ to $[0,1]$, meaning that $\mu$ satisfies three properties: (i) $\mu(T,S)\in [0,1]$ for all non-empty $T\subseteq S$ and for all $S\in \mathcal{X}$, (ii) $\sum_{T:T\subseteq S, T\ne \emptyset}\mu(T,S)=1$ for all $S\in \mathcal{X}$, and (iii) $\mu(T,S)=0$ whenever $T\not\subseteq S$. In this definition, $\mu(T,S)$ represents the probability of choosing a non-empty collection $T$ from choice set $S$.\footnote{Several models of $\mu$ examined in the paper allow the chosen collection to be empty; the decision maker is often assumed to select a default option in this case. Appendix \ref{sec: empty choices} provides characterizations of these models when $T$ can be empty.} We say that $\mu$ has full support if $\mu(T,S)>0$ for all pairs $(T,S)$ such that $T\subseteq S\subseteq X$ and $T$ is non-empty.

\subsection{Two common behavioral properties}\label{sec:properties}\label{sec:powerset}
We introduce two behavioral postulates that organize the class of models studied in the paper. The first axiom, Independence of Irrelevant Sets (IIS), adapts the Independence of Irrelevant Alternatives (IIA) axiom from choices over alternatives to choices over collections. IIS is formalized in Axiom \ref{ax: IIS} and states that the relative probability of choosing two collections is independent of the presence or absence of other collections. IIS therefore preserves the odds of collections that remain feasible. When one interprets each collection $T$ as a single composite alternative and each menu $S$ as a composite menu consisting of all nonempty subsets of $S$, $\mu(T,S)$ can be interpreted as the probability of choosing $T$ from the non-empty subsets of $S$. Under this interpretation, IIS intuitively follows from IIA.
\begin{ax}[Independence of Irrelevant Sets - IIS]\label{ax: IIS}
\[
\frac{\mu(T,S)}{\mu(T',S)}=\frac{\mu(T,S')}{\mu(T',S')} \quad \text{ for all non-empty } T,T' \subseteq S\cap S'
\]
as long as the two ratios are well defined and positive.
\end{ax}

The second axiom, Relative Additivity, links choice probabilities across two menus of different sizes and imposes a restriction on these probabilities when the menu changes.

\begin{ax}[Relative Additivity]\label{ax: RCG} For all $x,T,T'$, and $S$ such that $x\in S$ and $T,T' \subseteq S\setminus x$ and $T,T'\ne \emptyset$,
\[
\mu(T,S\setminus x) [\mu(T',S)+\mu(T'\cup x, S)] = \mu(T',S\setminus x) [\mu(T,S)+\mu(T\cup x, S)].
\]
\end{ax}

To understand Relative Additivity, consider some of its implications. First, when $\mu$ has full support, Relative Additivity states that
\[
 \frac{\mu(T,S\setminus x)}{\mu(T,S)+\mu(T\cup x, S)}=\frac{\mu(T',S\setminus x)}{\mu(T',S)+\mu(T'\cup x, S)} \quad \text{ for all non-empty $T,T'\subseteq S\setminus x.$}
\]
Consequently, the ratio $\frac{\mu(T,S\setminus x)}{\mu(T,S)+\mu(T\cup x, S)}$ is independent of $T$. To understand this ratio, suppose the initial choice set is $S\setminus x$ and $T$ is the chosen collection. Adding a new item $x$ to $S\setminus x$ may impact $T$ in two distinct ways. When $x$ has an indirect effect, the chosen collection remains the same and $T$ is selected with updated probability $\mu(T,S)$. Alternatively, $x$ may have a direct effect and be added to the chosen collection, in which case $T\cup x$ is selected with probability $\mu(T\cup x,S)$. Thus, the sum $\mu(T,S)+\mu(T\cup x, S)$ can be seen as the total choice probability related to $T$ under the expanded choice set. Hence, the ratio $\frac{\mu(T,S\setminus x)}{\mu(T,S)+\mu(T\cup x, S)}$ captures the relative probability of choosing a collection before and after adding the new option. Relative Additivity states that this relative probability is independent of the collection chosen initially. The multiplicative form of Relative Additivity in Axiom \ref{ax: RCG} accounts for the possibility that the ratio $\frac{\mu(T,S\setminus x)}{\mu(T,S)+\mu(T\cup x, S)}$ might not be well defined when $\mu$ does not have full support.

Second, Relative Additivity highlights a key distinction between the standard analysis of choices among alternatives and that of choices over collections. In the former, removing an alternative $x$ leaves all other options feasible, as alternatives are treated as separate objects. In the latter, removing $x$ makes every collection containing $x$ infeasible; one must therefore consider how these collections project onto the smaller menu. Relative Additivity accounts for this projection: for any nonempty $T\subseteq S\setminus x$, both $T$ and $T\cup x$ project to $T$ when $x$ is removed from $S\ni x$, explaining why the axiom involves the three terms $\mu(T,S\setminus x)$, $\mu(T,S)$, and $\mu(T\cup x,S)$. Consequently, unlike IIS, Relative Additivity does not admit the same direct analogy with standard axioms for choices among individual alternatives, even when each collection $T$ is treated as a composite alternative and $\mu(T,S)$ is interpreted as its choice probability from the power set of $S$.

Third, Relative Additivity implies that adding a new alternative to the choice set (weakly) decreases the choice probability of an existing collection. That is, $\mu(T,S)\le \mu(T,S\setminus x)$ for all non-empty $T\subseteq S\setminus x$ with $x\in S$. This is the monotonicity condition of $\mu$ studied in \cite{cattaneo2020random}. Consequently, Relative Additivity is sufficient for monotonicity.

Finally, when both denominators are positive, Relative Additivity implies
\[
 \frac{\mu(T,S\setminus x)}{\mu(T',S\setminus x)}=\frac{\mu(T,S)+\mu(T\cup x, S)}{\mu(T',S)+\mu(T'\cup x, S)} \quad \text{ for all non-empty $T,T'\subseteq S\setminus x$}.
\]
This formulation is reminiscent of the Proportionality property in \cite{tversky1972choice}.

\subsection{Models and characterizations}
In this section, we present each model and its characterization and show how the two behavioral postulates help organize these models into related classes and establish connections among them.

\subsubsection{Logit}\label{sec:LG}
\newcommand{\sx}{\sum_{T':\emptyset \ne T'\subseteq X}}
\newcommand{\sz}{\sum_{T':\emptyset \ne T'\subseteq S}}

In \cite{Brady-Rehbeck_2016_ECMA}, each collection $T$ is associated with a numerical value denoted by $\pi(T)$, where $\pi$ is a mapping from $\mathcal{X}$ to $\mathbb{R}_{++}$. \cite{Brady-Rehbeck_2016_ECMA} interpret $\pi(T)$ as the likelihood that $T$ is feasible in $X$. Depending on the context, the number $\pi(T)$ may represent the value, importance, salience, or weight of $T$ from the perspective of the decision maker (hereafter DM). For instance, in the product recommendation example, $\pi(T)$ may correspond to the platform's benefit from recommending a set $T$ of products to its customers. In the case of supermarket assortment choice, $\pi(T)$ may denote the sales revenue generated by items in $T$. Given a choice set $S$, the probability of choosing a non-empty $T\subseteq S$ is proportional to its value according to a logit formula: 
\[
\mu_{\texttt{LG}}(T,S)=\frac{\pi(T)}{\sum_{T':T'\subseteq S,T'\ne \emptyset} \pi (T')} \quad \text{ for all non-empty } T\subseteq S.
\]
The logit formulation of $\mu_{\texttt{LG}}$ is reminiscent of the Luce model in probabilistic choice theory \citep{luce1959individual}, with collections serving as composite alternatives. Consistent with the analogy between IIS and IIA that characterizes the Luce model, Proposition \ref{prop: logit} indicates that IIS fully characterizes $\mu_{\texttt{LG}}$.

\begin{prop}[Characterization of Logit $\mu$] \label{prop: logit} A full-support $\mu$ has a Logit representation if and only if it satisfies IIS.
\end{prop}

Proposition \ref{prop: logit} states that the logit formulation of $\mu$ in \cite{Brady-Rehbeck_2016_ECMA} is captured by a simple behavioral postulate. The necessity part in Proposition \ref{prop: logit} is straightforward. For the sufficiency direction, the proof is constructive and follows an approach similar to that used in the characterization of the Luce model.

\subsubsection{Random Categorization}\label{sec:RCG}
Consider a decision maker who divides the grand set into different categories. Each category represents a bundle of alternatives. The categories can overlap with each other. The DM selects a particular category and subsequently chooses a collection by including all available options that belong to the category she considers. In simple terms, the chosen collection is the intersection of the choice set and the considered category. \cite{Aguiar_2017_EL} calls this model Random Categorization (RCG).

Formally, \citet{Aguiar_2017_EL} assumes that each category $C$ is a subset of the grand set $X$. Each category $C$ has an associated probability $m(C)$, where $m\colon \mathcal{X}\to [0,1]$ is a probability distribution over $\mathcal{X}$.\footnote{\citet{Aguiar_2017_EL} assumes that the collection of categories is a strict subset of $\mathcal{X}$. However, one can easily extend the collection of categories to $\mathcal{X}$ by simply assigning zero weight to all non-empty subsets of $X$ that do not belong to the original set of categories.} The decision maker first considers a category $C$, which happens with probability $m(C)$. If there is nothing for the decision maker to choose from, i.e., $C\cap S=\emptyset$, the DM redraws another category until finding one category $C'$ such that $C'\cap S\ne \emptyset$. For this process to terminate almost surely after finitely many draws, each option $x\in X$ is assumed to belong to at least one category $C$ with $m(C)>0$. The probability of choosing a non-empty $T\subseteq S$ is the sum over all categories $C$ for which the intersection of $S$ and $C$ equals $T$, normalized by the total probability of all categories overlapping with $S$: 
\[
 \mu_{\texttt{RCG}}(T,S)=\frac{\sum_{C}m(C)\mathbbm{1}(T=C\cap S)}{\sum_{C: S\cap C\ne\emptyset} m(C)} \quad \text{ for all non-empty } T\subseteq S.
\]
Note that the normalization term, $\sum_{C: S\cap C\ne\emptyset} m(C)$, arises from the non-emptiness of the chosen collection. When the chosen collection is allowed to be empty, this term disappears.\footnote{Appendix \ref{sec: empty choices} provides a characterization of $\mu_{\texttt{RCG}}$ when the chosen collection can be empty.}

The RCG model is applicable to the assortment choice of supermarkets or product recommendations by platforms mentioned earlier. For instance, in the supermarket assortment choice example, each category $C$ may represent a product category (fruits, vegetables, etc.), and $m(C)$ may correspond to the probability that items in $C$ are in stock. The supermarket would like to offer items in $C$ to customers, but only items in $S$ are available; hence, the supermarket offers everything in $C\cap S$. In the product recommendation example, each category $C$ may also represent a product category.

The RCG model satisfies Relative Additivity. In this model, Relative Additivity arises from a separation between what is chosen (collection $T$) and how its probability is assigned (via category $C$). Specifically, the chosen collection is simply an outcome of the category selection. The probability of choosing $T$ depends entirely on the probability distribution over categories and is independent of any properties of $T$. Once a category is drawn, there is no additional weight applied directly to the chosen set. This independence structure gives rise to Relative Additivity. More generally, Relative Additivity emerges in models that exhibit this independence structure and disappears when the structure is absent (see Section \ref{sec:salience} for details).

The RCG model additionally satisfies Positivity-1, which concerns the coverage of chosen collections.

\begin{ax}[Positivity-1]\label{ax: RCG-Positivity} For all $(x,S)\in X\times\mathcal{X}$ with $x\in S$, there exists $T\ni x$ such that $\mu(T,S)>0$. 
\end{ax}

Positivity-1 arises from the assumption that each option $x$ belongs to at least one category $C$ with $m(C)>0$. To elaborate, suppose $x\in S$. Let $C$ be an arbitrary category that includes $x$ with $m(C)>0$. Then $T=C\cap S$ is chosen with positive probability from choice set $S$. Clearly, $T$ contains $x$. Note that Positivity-1 is trivially satisfied when $\mu$ has full support.

We are now ready to state a characterization of RCG. Proposition \ref{prop: RCG} establishes that the RCG model is characterized by Axioms \ref{ax: RCG}-\ref{ax: RCG-Positivity}.

\begin{prop}[Characterization of RCG] \label{prop: RCG} $\mu$ has an RCG representation if and only if it satisfies Positivity-1 and Relative Additivity.
\end{prop}

Proposition \ref{prop: RCG} allows the researcher to test and falsify the RCG model by checking two simple conditions. The proof of Proposition \ref{prop: RCG} is constructive and uses induction based on the number of alternatives in the choice set. First, for each category $C\in \mathcal{X}$, we define its probability as $m(C)=\mu(C,X)$. By construction, $m$ is a probability distribution over categories, and $\mu$ has an RCG representation at $X$. Second, we use Relative Additivity to show that $\mu$ has an RCG representation at choice sets having $|X|-1,\dots, 2$ elements by induction. The induction step utilizes the fact that Relative Additivity connects choice probabilities at two choice sets of different sizes ($S$ and $S\setminus x$), so we can apply the representation at a larger choice set to obtain the representation in a smaller choice set.

\subsubsection{Independent Choice}

In \cite{Manzini-Mariotti_2014_ECMA}, the choice probability of a collection depends on the probabilities of choosing options within the collection. Each option $x\in X$ has an independent parameter $\gamma(x)\in (0,1)$, which represents the probability of that option being selected. In the supermarket assortment example, $\gamma(x)$ may correspond to the probability that product $x$ is in stock. In the product-recommendation setting, $\gamma(x)$ may denote the probability that a similar customer buys $x$.

For a given choice set $S$, a non-empty subset $T$ of $S$ is chosen if and only if the DM selects every alternative in $T$ while ignoring all options in $S\setminus T$. The probability of choosing $T$ is defined as
\[
\mu_{\texttt{IC}}(T,S) =
      \frac{\prod\limits_{x\in T} \gamma(x) \prod\limits_{y\in S\setminus T} (1-\gamma(y))}{1-\prod_{z\in S}(1-\gamma(z))}  \quad \text{ for all non-empty } T\subseteq S.
\]
In the formulation of $\mu_{\texttt{IC}}(T,S)$, the normalization term, $1 - \prod_{z \in S} (1 - \gamma(z))$, represents the probability of drawing a non-empty subset of $S$. It arises from the non-emptiness of the chosen collection. In the decision-making process, this normalization ensures that if the DM draws an empty collection as her choice, which occurs with probability $\prod_{z \in S} [1 - \gamma(z)]$, she will redraw another collection (with replacement) until a non-empty collection is selected.\footnote{When the chosen collection is allowed to be empty, the normalization term disappears. Appendix \ref{sec: empty choices} provides a characterization of $\mu_{\texttt{IC}}$ in this case. \cite{regenwetter1998random} give an alternative characterization of this model. See Section \ref{sec: Literature Review} for further discussion of the connection between their results and ours.}

The three formulations of $\mu$ introduced, Logit, Random Categorization, and Independent Choice, differ in their functional forms and underlying narratives. One may wonder whether the three models are related in a particular way. Interestingly, there is a notable connection between them: Independent Choice is equivalent to the intersection of Logit and Random Categorization. In Proposition \ref{prop: IC} below, we present a behavioral characterization of Independent Choice (IC). We show that a full-support $\mu$ has an Independent Choice representation if and only if it satisfies IIS and Relative Additivity. Consequently, the relationship between the three models follows directly from their behavioral foundations.

\begin{prop}[Characterization of IC]\label{prop: IC} A full-support $\mu$ has an Independent Choice representation if and only if it satisfies IIS and Relative Additivity.
\end{prop}

Proposition \ref{prop: IC} indicates that verifying IIS and Relative Additivity is sufficient to determine whether a full-support $\mu$ has an Independent Choice representation. Proposition \ref{prop: IC}, together with Propositions \ref{prop: logit}-\ref{prop: RCG}, sheds light on the connection between Logit, Random Categorization, and Independent Choice. Note that the relationship among the three models is not new and was also documented in \cite{kovach2023reference} in their study of reference-dependent probabilistic choice in a slightly different setting.\footnote{More precisely, they document the relationship between the three models when the chosen collection is allowed to be empty. Our result here applies to the case when the selected collection must be non-empty. In Appendix \ref{sec: empty choices}, we provide characterizations of RCG and IC when the DM can select an empty set. Through the behavioral characterizations, we also re-establish their result.} \cite{kovach2023reference} do not investigate the behavioral implications of the three models and use their functional forms to identify the relationship. We employ an alternative approach by using their behavioral foundations. Our axiomatic approach clarifies the implications of the functional form assumptions and guides the analyst in selecting the most suitable model for their application of interest. Moreover, the relationships among different models of $\mu$ often become evident through their behavioral characterizations. As we will show in Section \ref{sec: relationship}, our approach is particularly useful for understanding the connections between various models of $\mu$ when they exhibit more complex functional forms and are motivated by highly distinct narratives.

The proof of sufficiency in Proposition \ref{prop: IC} proceeds as follows. First, we use the IIS axiom and apply Proposition \ref{prop: logit} to obtain a Logit representation of $\mu$, allowing us to express $\mu(T,S)=\frac{\pi(T)}{\sum_{T':T'\subseteq S,T'\ne \emptyset} \pi (T')}$ for some $\pi\colon \mathcal{X}\rightarrow \mathbb{R}_{++}$. Next, we apply Relative Additivity to this functional form and, through induction, demonstrate that the Logit representation can be rewritten as an Independent Choice representation by defining $\gamma(x)=\frac{\pi(X)}{\pi(X)+\pi(X\setminus x)}$ for all options $x$. 

The proof of Proposition \ref{prop: IC} also sheds light on the relationship among the primitives in the Logit, Random Categorization, and Independent Choice models. Specifically, given the normalization $\sum_{T':\emptyset \ne T'\subseteq X} \pi (T')=1$ in the Logit model, we show that the probability measure over categories in the RCG representation must correspond to the $\pi$ function in the Logit representation: $m(C)=\pi(C)$ for all $C\in \mathcal{X}$. Consequently, the equations $\gamma(x)=\frac{\pi(X)}{\pi(X)+\pi(X\setminus x)}$ for all $x\in X$ and $m(C)=\pi(C)$ for all $C\in \mathcal{X}$ jointly characterize the interconnection between the primitives in the three models.


\subsubsection{Salience Models}\label{sec:salience}
In this section, we investigate three models that incorporate the salience of various aspects of the choice environment into the decision-making process. These models differ in the type of salience they consider: (i) salience of attributes in Elimination By Aspects \citep{Tversky:1972}, (ii) salience of alternatives in Random Reference Model \citep{kibris2024random}, and (iii) salience of nests in Nested Stochastic Collection \citep{kovach2022behavioral}.

\medskip
\noindent \textbf{Salience of Attributes.} Many products in real life have observable attributes. For instance, a typical soft drink can be differentiated from others by its price, size, design, calorie content, and intended use. Consider a DM who pays attention to a particular attribute and ignores options that do not possess that attribute when making choices. To illustrate, when recommending soft drinks on hot days, a vending machine may prioritize refreshing and hydrating options, such as sparkling water and lemonade, while neglecting heavier soft drinks like colas and sodas. \cite{Tversky:1972} calls this model Elimination By Aspects (EBA). Throughout the paper, we focus on the so-called static version of EBA, which is characterized by a one-time elimination of alternatives, as the original EBA model developed by \cite{Tversky:1972} allows for a sequence of eliminations until only one option remains. This focus allows the decision makers to choose a collection of options and facilitates comparisons across models, as all other formulations of $\mu$ examined in this paper are also static. With a slight abuse of terminology, we will refer to the static EBA model simply as EBA. When clarification is necessary, we will refer to the original EBA model as sequential EBA.

Formally, let $\{i, j, k,\dots\}$ denote the finite set of all observable attributes. Let $\omega_i> 0$ be the level of attention that attribute $i$ receives. It measures the salience (or value) of aspect $i$. Without loss of generality, assume that $\sum_i \omega_i=1$. Let $A_i\ne \emptyset$ be the set of options having aspect $i$. Each option has at least one attribute. Suppose aspect $i$ is salient. All options without aspect $i$ fail to receive attention. The DM chooses all feasible options having aspect $i$; hence, the selected collection is $A_i \cap S$, with $S$ being the choice set. When $A_i \cap S=\emptyset$, the DM switches attention to a new attribute until finding some aspect $j$ such that $A_j \cap S\ne \emptyset$. The probability of choosing a non-empty $T\subseteq S$ is the normalized sum over all attributes $i$ satisfying $T=A_i\cap S$:\footnote{In Appendix \ref{sec: Attribute Rule}, we show that the static EBA model described here is identical to the first stage of the choice procedure (after reformulation) in the Attribute Rule model introduced by \cite{Gul_Natenzon_Pesendorfer_2014_ECMA}. Our equivalence result sheds light on the differences in choice behaviors between the original EBA model and the Attribute Rule model noted by \cite{Gul_Natenzon_Pesendorfer_2014_ECMA}. See Appendix \ref{sec: Attribute Rule} for details.}

\begin{equation}\label{eq: EBA1}
   \mu_{\texttt{EBA}}(T,S)=\sum_{i}\frac{\omega_i \mathbbm{1}(T=A_i\cap S)}{\sum_{j:A_j\cap S\ne \emptyset}\omega_j}  \quad \text{ for all non-empty } T\subseteq S.
\end{equation}

In the supermarket assortment choice example, aspect $i$ may represent the shelf life of a product. The collection $A_i$ could then denote the group of items with similar shelf life. As only products in $S$ are available, the supermarket offers items in $A_i\cap S$ to its customers. 

To provide a characterization of EBA, we first document a connection between EBA and RCG studied in Section \ref{sec:RCG}. In Remark \ref{rem:EBA-RCG}, we demonstrate that EBA is nested in RCG by showing that one can obtain a representation for the latter from a representation of the former by appropriately defining the probability distribution over the categories. 

\begin{rem}\label{rem:EBA-RCG} Let the distribution over categories $m\colon\mathcal{X}\to [0,1]$ in the RCG model be such that $m(C)= \sum_{i: A_i=C}\omega_i$. For all non-empty $S\subseteq X$ and all non-empty $T\subseteq S$, 
\[
   \mu_{\texttt{EBA}}(T,S)=\mu_{\texttt{RCG}}(T,S).
\]
\end{rem}
Remark \ref{rem:EBA-RCG} follows from the observation that each category $C$ in the RCG model can be interpreted as the set of options possessing a specific attribute in the EBA model. If $m(C)=\sum_{i: A_i=C}\omega_i$, i.e., the probability of drawing category $C$ equals the total consideration probability of all attributes $i$ whose collection of options having aspect $i$ is exactly $C$, then the two models generate identical choice behavior.

Given that RCG nests EBA as a particular case, it follows from the characterization of $\mu_{\texttt{RCG}}$ in Proposition \ref{prop: RCG} that Relative Additivity (Axiom \ref{ax: RCG}) and Positivity-1 (Axiom \ref{ax: RCG-Positivity}) are necessary for the EBA model. The two axioms, however, are not sufficient for a characterization of EBA. Note that in the EBA model, the DM first selects a particular attribute and subsequently chooses a collection containing all feasible options possessing that attribute. Hence, the decision maker's choices depend on the set of attributes, which is assumed to be observable and exogenous. This exogeneity assumption of attributes imposes a stronger restriction on $\mu$ than Positivity-1. We refer to this stronger condition as Positivity-2 (Axiom \ref{ax: EBA-posi}). Note that Positivity-2 is nested in Positivity-1, as the former implies the latter.

\begin{ax}[Positivity-2]\label{ax: EBA-posi} For all non-empty $T\subseteq S$, $\mu(T,S)>0$ if and only if there exists an attribute $i$ such that $T=A_i\cap S$. 
\end{ax}

Positivity-2 follows from the definition of the EBA model. Positivity-2 states that the DM selects a collection with positive probability if and only if it corresponds to the set of all feasible options possessing a specific attribute. By combining Positivity-2 and Relative Additivity, we obtain a characterization of EBA. This is established in Proposition \ref{prop: EBA}.

\begin{prop}[Characterization of EBA] \label{prop: EBA}$\mu$ has an EBA representation if and only if it satisfies Positivity-2 and Relative Additivity.
\end{prop}

Proposition \ref{prop: EBA} allows the researcher to verify whether the decision maker's choice aligns with the Elimination by Aspects model. The proof of Proposition \ref{prop: EBA} follows from the characterization of $\mu_{\texttt{RCG}}$ in Proposition \ref{prop: RCG}.
 
We complete this section by commenting on the exogeneity assumption of attributes in EBA. The model assumes that attributes are exogenously given and observed by the outside analyst. Such an assumption can be overly restrictive in certain contexts, as the outside analyst may not directly observe the attributes. Hence, relaxing the exogeneity assumption by endogenizing attributes can improve the practical relevance of the model. Additionally, it allows us to compare EBA to other models that do not have attributes in the choice environment.

To investigate the EBA model when attributes are endogenous, we first define an endogenous EBA model in Definition \ref{def: endogenous EBA}. 

\begin{defn}[Endogenous EBA]\label{def: endogenous EBA} $\mu$ has an endogenous EBA representation if there exists a finite collection of attributes $\{i,j,k,\dots\}$, with each option having at least one attribute, such that $\mu$ has an EBA representation with $\{i,j,k,\dots\}$ being the set of attributes.  
\end{defn}

Note that as the attributes in the ordinary sense are not observed, the term \textit{attributes} in Definition \ref{def: endogenous EBA} does not possess a specific meaning beyond functioning as a label. In particular, the outside analyst can define \textit{artificial aspects} $\{i,j,k,\dots\}$ and the associated set of options $A_i$ possessing attribute $i$ to test whether the dataset adheres to the elimination of the defined artificial aspects.

In environments where attributes are endogenous, Remark \ref{rem: EBA-AR-RCG} shows that EBA and RCG are equivalent. Note that when attributes are observable, Remark \ref{rem:EBA-RCG} states that EBA is strictly nested in RCG, with the strictness coming from the exogeneity of attributes. Endogenizing the attributes eliminates the condition imposed by their exogeneity; hence, EBA corresponds to RCG without this additional restriction.

\begin{rem}\label{rem: EBA-AR-RCG} $\mu$ has an endogenous (static) EBA representation if and only if it has an RCG representation. 
\end{rem}

Together with Proposition \ref{prop: RCG}, Remark \ref{rem: EBA-AR-RCG} implies that verifying Positivity-1 and Relative Additivity is sufficient to determine whether the DM's choice adheres to the one-time elimination of aspects, where the aspects can be defined arbitrarily by the analyst.


\medskip
\noindent \textbf{Salience of Alternatives.} Many studies in psychology and economics have documented that the decision maker's behaviors often depend on what she perceives as a reference point  \citep{samuelson1988status,Knetsch89,kahneman1992reference}.\footnote{Reference points can be default options, initial endowments, entitlements, or past choices.} There are two main views on modeling these reference-dependent behaviors. The first view assumes that the reference point directly impacts the DM's evaluation and explicitly incorporates the reference point into the DM's utility function. In the second view, the reference point does not directly affect the DM's utility function but influences what the DM perceives and is willing to select.\footnote{Note that the two views are not mutually exclusive.} The Random Reference Model (RRM) in this section adopts the second view above and is based on \cite{masatlioglu2014canonical} and \cite{kibris2024random}.

There exist multiple reference points. The reference point is endogenously determined by a random process. The RRM restricts the reference point to be an option within the choice set, i.e., a reference point must be a feasible alternative. Each option $x\in X$ has an associated value $s_x>0$, which may represent its salience, value, or importance. The probability of an option being the reference point depends on its salience relative to the total salience of all feasible alternatives. Given choice set $S$, option $x\in S$ is the reference point with probability $\frac{s_x}{\sum_{y\in S} s_y}$. When alternative $x$ serves as the reference point, it imposes a constraint on the DM through a constraint set $\mathcal{Q}(x)\subseteq X$. The set $\mathcal{Q}(x)$ is interpreted as the collection of options that are appealing from the perspective of $x$. It may correspond to the set of options that $x$ reminds the decision maker of, the set of options that are similar to $x$, or the set of options that are unambiguously better than $x$. Options that are not appealing from the perspective of $x$ fail to receive attention. The DM selects all feasible alternatives within her constraint set. Given choice set $S$, the probability of choosing a non-empty $T\subseteq S$ is defined as
\begin{equation}\label{eq: RRM}
\mu_{\texttt{RRM}}(T,S)=\sum_{x\in S} \frac{s_x}{\sum_{y\in S} s_y}\mathbbm{1}(T=\mathcal{Q}(x)\cap S) \quad \text{ for all non-empty } T\subseteq S.
\end{equation}

In the supermarket assortment choice example, the reference point $x$ may represent a product currently in high demand, and the constraint set associated with $x$, $\mathcal{Q}(x)$, may include complementary items or products frequently purchased together with $x$. In the case of Japanese vending machines, as the machine can scan the current customer's face and recall similar past customers from its database, the reference point $x$ may represent the most similar past purchase. The set $\mathcal{Q}(x)$ may then represent the collection of drinks similar to $x$.

Regarding the constraint set, \cite{masatlioglu2014canonical} assume that each option must be appealing from the perspective of itself, i.e., each alternative $x$ is always included in its constraint set $\mathcal{Q}(x)$. This property is related to experimental findings that one alternative becomes more desirable when it serves as the reference point \citep{Knetsch89}. We further assume that the constraint sets associated with two different options are distinct: $\mathcal{Q}(x)\ne \mathcal{Q}(y)$ when $x\ne y$. This assumption allows the decision maker to possibly make different choices under two reference points, even when they are equally salient. The property is related to an experimental finding that individuals often exhibit distinct behaviors under different reference points \citep{sullivan1995effect,lin2006multiple}.

The functional form of $\mu_{\texttt{RRM}}$ resembles that in both EBA and RCG. All three models formulate the decision maker's choice as a collection of feasible options satisfying certain properties. In EBA, these options possess the salient attribute. In RCG, they fall into the same category. In RRM, they belong to a constraint set imposed by the salient alternative (the reference point). Given the similarities in how the three models formalize choices, as we will show, all behavioral postulates in RRM are connected to those in EBA and RCG. However, the conditions imposed on the constraint sets and the feasibility of the reference point distinguish RRM from EBA and RCG.\footnote{In Section \ref{sec: relationship}, we show that RRM and RCG intersect only in a particular case where the decision maker always selects singleton sets.}

One of the key axioms in the characterization of RRM (see Appendix \ref{appendix:RRM_characterization} for details of the characterization) is a generalization of Relative Additivity in EBA and RCG models. To introduce this axiom, define the \textit{revealed constraint function $\mathcal{Q}^R\colon X\rightarrow \mathcal{X}$} as follows. For each $x\in X$, $\mathcal{Q}^R(x)$ includes $x$ and all options $y\ne x$ for which $\mu(\{x\},\xy)=0$. Mathematically,
\[
\mathcal{Q}^R(x)=\{x\}\cup \{y: y\in X, y\ne x \, \text{ such that } \, \mu(\{x\},\xy)=0\} \quad \text{ for all } x\in X.
\]
The revealed constraint function $\mathcal{Q}^R$ corresponds to the underlying constraint function if $\mu$ has an RRM representation. The main axiom in the characterization of RRM reads as follows.

\begin{ax}[Relative Additivity-1]\label{ax: RRM1} For all non-empty $T,T'\subseteq S\setminus x$ and $x\in S$ such that $T,T'\ne \mathcal{Q}^R(x)\cap (S\setminus x)$ $$\mu(T,S\setminus x) [\mu(T',S)+\mu(T'\cup x, S)]=\mu(T',S\setminus x) [\mu(T,S)+\mu(T\cup x, S)].$$
\end{ax}
Axiom \ref{ax: RRM1} generalizes the original Relative Additivity by narrowing down its applicability through explicitly requiring $T,T'$ to satisfy an additional condition: $T,T'\ne \mathcal{Q}^R(x)\cap (S\setminus x)$. This additional constraint arises from the assumption that the reference point must be feasible and included in the associated constraint set. Specifically, the RRM allows $x$ to be a reference point under choice set $S\ni x$, but it cannot be the reference point under $S\setminus x$ because $x$ is no longer available. To understand how this assumption affects Relative Additivity, consider a scenario where $T=\mathcal{Q}^R(x) \cap (S\setminus x)$, so the condition in Axiom \ref{ax: RRM1} fails to hold. It follows that $T\cup x=\mathcal{Q}^R(x) \cap S$ because $x\in \mathcal{Q}^R(x)$. Consequently, the reference point $x$ influences the probability of choosing $T\cup x$ under $S$. However, there are no similar impacts under the choice set $S\setminus x$ because $x$ cannot serve as a reference point when it is no longer feasible. Hence, Relative Additivity need not hold, and that explains why the condition $T,T'\ne \mathcal{Q}^R(x)\cap (S\setminus x)$ is present in Axiom \ref{ax: RRM1}.


\medskip
\noindent \textbf{Salience of Nests.} Nested logit is one of the most common parametric frameworks for analyzing probabilistic choice. \cite{kovach2022behavioral} generalize the nested logit model by introducing a Nested Stochastic Collection (NSC).\footnote{See Appendix \ref{appendix:nested_logit} for the relationship among nested logit and other models in our framework.} In an NSC, there exist pairwise disjoint nests $N_1,N_2,\dots, N_q$ ($q\ge 1$) that partition the grand set. Each nest represents a collection of alternatives sharing some common characteristics. Every nest has an associated value. The DM first draws a salient nest and subsequently selects a collection comprising all feasible options within the salient nest. The DM's choice over nests is governed by a Luce formula through a weighting (or salience) function $\sigma\colon \mathcal{X}\cup \{\emptyset\} \rightarrow \mathbb{R}_{+}$. The function $\sigma$ assigns a weight $\sigma(T)$ to each collection $T$, where the weight $\sigma(T)$ may represent the salience, value, or importance of $T$. \cite{kovach2022behavioral} further assume that $\sigma(\emptyset)=0$ and $\sigma(T)>0$ whenever $T\ne \emptyset$. Given choice set $S$, the probability of choosing a non-empty $T\subseteq S$ is defined as
\begin{equation}\label{eq: NSC}
\mu_{\texttt{NSC}}(T,S)=\sum_{i=1}^q\frac{\sigma(N_i\cap S)}{\sum_{j=1}^q \sigma(N_j\cap S)}\mathbbm{1}(T=N_i\cap S) \quad \text{ for all non-empty } T\subseteq S.
\end{equation}

In the supermarket assortment choice example, each nest may represent a distinct product category, such as meat, fish, frozen vegetables, or dairy, while $\sigma(T)$ may indicate the sales of the items in $T$. In the Japanese vending machine context, each nest may correspond to a specific product brand, such as Coke or 7UP, with $\sigma(T)$ denoting the within-day revenue of drinks in $T$.

The formulation of $\mu$ in equation (\ref{eq: NSC}) looks similar to that in EBA, RCG, or RRM. As each nest in NSC is a collection of alternatives, the notion of a salient nest in NSC is conceptually comparable to the notions of a category in RCG, a set of options possessing a specific attribute in EBA, or a constraint set in RRM. The primary distinction between NSC and these three models lies in how the weighting functions impact choice probabilities. In NSC, the probability of choosing a collection depends on the collection's salience value, as $\sigma(T)$ explicitly appears in $\mu_{\texttt{NSC}}(T,S)$. In contrast, in EBA, RCG, or RRM, the probability of selecting a collection is independent of its salience, properties, or structure. Instead, this probability depends on category-specific weight (in RCG), attribute-specific weight (in EBA), or item-specific weight (in RRM). More generally, EBA, RCG, and RRM exhibit an independence structure between what is chosen and how its probability is assigned. Meanwhile, in NSC, such an independence structure is absent. As we will show in Section \ref{sec: relationship}, this fundamental distinction sets NSC apart from the other three models.

Because of the way in which the probability of choosing a collection is assigned, NSC is behaviorally closest to the Logit model of \cite{Brady-Rehbeck_2016_ECMA}. Intuitively, this similarity arises from the fact that \cite{Brady-Rehbeck_2016_ECMA} adopt a logit functional form, whereas NSC is a generalization of the nested logit. In Section \ref{sec:LG}, we showed that a full-support Logit $\mu$ is characterized by IIS (Axiom \ref{ax: IIS}). The primary behavioral postulate in NSC, Path-Independence of Irrelevant Sets (PIIS, Axiom \ref{ax: NSC-PIIS}), is conceptually analogous to IIS.\footnote{See Appendix \ref{appendix:NSC_characterization} for a complete characterization of NSC.} Broadly speaking, PIIS states that $\frac{\mu(T,S)}{\mu(T^*,S)}\frac{\mu(T^*,S')}{\mu(T',S')}$ is independent of the collection $T^*$ (the ``path") and choice sets $S,S'$ whenever the two ratios are well defined and positive.

\begin{ax}[Path-Independence of Irrelevant Sets, PIIS]\label{ax: NSC-PIIS} For arbitrary $T$ and $T'$,
\[
\frac{\mu(T,S)}{\mu(T^*,S)}\frac{\mu(T^*,S')}{\mu(T',S')} \quad \text{ is independent of the choice of $(T^*,S,S')$}
\]
as long as the two ratios are well defined and positive.
\end{ax}

PIIS constitutes a generalization of IIS. To elaborate, whenever two collections $T$ and $T'$ are selected with strictly positive probabilities in both choice sets $S$ and $S'$, applying PIIS with $(T^*,S,S')=(T,S,S')$ and $(T^*,S,S')=(T',S,S')$ yields IIS. This scenario, however, does not always occur in the NSC model because the decision maker may choose certain collections with zero probability. Consequently, PIIS can be interpreted as a generalization of IIS that imposes additional restrictions involving intermediate collections (the ``paths") when $\mu$ does not have full support. Under full support of $\mu$, PIIS and IIS are equivalent.

\bigskip
\section{Relationship between models}\label{sec: relationship}
In this section, we investigate the relationship between various models of $\mu$ studied in Section \ref{sec:models_and_characterization}. We first define three particular cases of $\mu$. In Definition \ref{defn:deterministic with full choice-mu}, we say that $\mu$ is deterministic-with-full-choice if the decision maker selects everything in the choice set with certainty. In the consideration set literature, this specification of $\mu$ is often referred to as full attention.

\begin{defn}\label{defn:deterministic with full choice-mu} $\mu$ is deterministic-with-full-choice if $\mu(S,S)=1$ for all non-empty $S\subseteq X$.
\end{defn}

In Definition \ref{defn: Singleton-mu} below, we introduce a singleton representation of $\mu$. Under a singleton $\mu$, the DM chooses a collection with positive probability if and only if it contains only one option (part (i)). Additionally, the relative choice probabilities are independent of the choice set (part (ii)). Note that the singleton $\mu$ and the Luce model in the probabilistic choice literature are distinct, as they operate in two different domains. However, one can obtain the Luce model from a singleton $\mu$ by appropriately defining choice probabilities.\footnote{To elaborate, suppose $\mu$ has a singleton representation. Let $p\colon X\times \mathcal{X}\to [0,1]$ be a stochastic choice function. For all non-empty $S\subseteq X$, define $p(x,S)=0$ if $x\not\in S$ and $p(x,S)=\mu(\{x\},S)$ if $x\in S$. It is straightforward to verify that $p$ has a Luce representation.}

\begin{defn}[Singleton-$\mu$]\label{defn: Singleton-mu} $\mu$ admits a singleton representation if 
\vspace{-0.5em}
\begin{enumerate}[label=(\roman*)]
\itemsep0em
    \item $\mu(\{x\},S)>0$ for all $x\in S$, for all $S\subseteq X,$ and $\mu(T,S)=0$ whenever $|T|\ge 2$;
    \item $\frac{\mu(\{x\},S)}{\mu(\{y\},S)}=\frac{\mu(\{x\},S')}{\mu(\{y\},S')}$ for all $S,S'\supseteq \xy$.
\end{enumerate}
\end{defn}

Definition \ref{defn: nest-invariant attention} below presents a particular case of the NSC model, called nest-invariant $\mu$, where the weighting function remains constant within the same nest: $\sigma(T)=\sigma(T')$ whenever $T,T'\ne \emptyset$ and $T,T'\subseteq N_i$ for some $i$. This scenario may correspond to an environment where every collection of items within the same nest is equally salient, or to a decision maker who values such collections equally.

\begin{defn}[Nest-invariant $\mu$]\label{defn: nest-invariant attention} $\mu$ has a nest-invariant representation if $\mu$ has an NSC representation with a nest-invariant weighting function. 
\end{defn}

The nest-invariant $\mu$ encompasses both the deterministic-with-full-choice $\mu$ and the singleton $\mu$ as special cases. Specifically, the deterministic-with-full-choice $\mu$ corresponds to a nest-invariant $\mu$ wherein there exists a single nest that is identical to the grand set. Meanwhile, the singleton $\mu$ is equivalent to a nest-invariant $\mu$ where there are $|X|$ nests, each consisting of a distinct option.

Although the nest-invariant $\mu$ is defined in terms of the model's unobserved primitive (the weighting function), it can also be characterized solely by the observed data. The nest-invariant $\mu$ satisfies a property called the probabilistic attention filter. This property specifies conditions under which adding a new item to the choice set does not affect the choice probabilities of a collection chosen initially. It is a generalization of the attention filter condition studied in \cite{Masatlioglu-Nakajima-Ozbay_2012_AER} to a probabilistic setting.\footnote{The deterministic attention filter condition states that choice is unaffected by an overlooked alternative: $\Gamma(S)=\Gamma(S\setminus x)$ if $x\in S\setminus \Gamma(S)$ and $x\in S$. Here, $\Gamma(S)$ denotes the (deterministic) chosen collection from choice set $S$.}

\begin{defn}[Probabilistic attention filter]\label{defn: probabilistic attention filter} Take $x\in S$ and non-empty $T\subseteq S\setminus x$. Then $\mu(T,S)=\mu(T,S\setminus x)$ if $\mu(\{x\},S)=0$ and $\mu(T,S)>0$.
\end{defn}
To understand the probabilistic attention filter property, consider a scenario where the initial choice set is $S\setminus x$ and $T\ne \emptyset$ is the chosen collection, which implies that $\mu(T,S\setminus x)>0$. The property states that adding $x$ to the choice set does not impact the choice probability of $T$ provided that two conditions are satisfied: (i) $\{x\}$ is not selected in the expanded choice set, and (ii) $T$ is still chosen with positive probability in the expanded choice set. Intuitively, these conditions ensure that the introduction of $x$ does not shift the decision maker's focus away from $T$; hence, the choice probability of $T$ remains unchanged. We will show in Proposition \ref{prop: NSC-relationship} that the nest-invariant $\mu$ is equivalent to an NSC that satisfies the probabilistic attention filter condition. Consequently, the nest-invariant $\mu$ is characterized by three axioms in the NSC model (see Appendix \ref{appendix:NSC_characterization}) along with the probabilistic attention filter property. 

We are now ready to provide the connections between different models of $\mu$. 

\smallskip
\noindent \textbf{Relationship between Logit, IC, and other models.} As mentioned earlier in the discussion following Proposition \ref{prop: IC}, the intersection of Logit and RCG is identical to IC. Additionally, Remark \ref{rem: EBA-AR-RCG} states that RCG and endogenous (static) EBA are equivalent. Therefore, the intersection of Logit and endogenous EBA is also identical to IC. This result implies that the intersection of two models of $\mu$ that have proven useful in capturing individuals' behavior has its own distinct appeal.

Regarding other formulations of $\mu$, note that Logit requires $\mu$ to have full support. In both RRM and NSC, this full-support condition fails to hold, as the DM always chooses some collections with zero probability. For instance, in NSC, two distinct and overlapping collections cannot be selected with positive probability from the same choice set. Consequently, Logit $\mu$ is disjoint from RRM and NSC. Additionally, since IC is nested in Logit following Propositions \ref{prop: logit}-\ref{prop: IC}, IC is also disjoint from RRM and NSC. These results are stated in Remark \ref{rem: BR-relationship}.

\begin{rem}[Logit, IC, and other models]\label{rem: BR-relationship} Suppose $\mu$ has an RRM or NSC representation. Then it does not have a Logit or an IC representation.
\end{rem}

\noindent \textbf{Relationship between RRM and other models.} Proposition \ref{prop: RRM-relationship} below summarizes the connections between RRM and other formulations of $\mu$. It establishes that RRM is sharply distinct from these models, as RRM intersects other models only in a particular case where $\mu$ has a singleton representation.

\begin{prop}[RRM and other models]\label{prop: RRM-relationship} The following are equivalent: 
\vspace{-1em}
\begin{enumerate}[label=(\roman*)]
\itemsep-0.5em
    \item $\mu$ has a singleton representation
    \item $\mu$ has both RRM and endogenous EBA (or RCG) representations
    \item $\mu$ has both RRM and NSC representations.
\end{enumerate}
\end{prop}

Two comments follow from Proposition \ref{prop: RRM-relationship}. First, although conceptually similar in terms of the underlying choice procedure and behavioral postulates, RRM intersects endogenous EBA (or RCG) only in a particular case when $\mu$ has a singleton representation. The strong distinction between RRM and these two models primarily comes from the restrictions imposed on the reference points and constraint sets in RRM. When these restrictions are relaxed, the intersection of these models enlarges.

Second, Proposition \ref{prop: RRM-relationship} implies that RRM and NSC intersect when the constraint sets in RRM and the collection of nests in NSC share a common structure. Specifically, in the RRM, each constraint set must consist of a single option. Similarly, each nest in the NSC model also comprises only one alternative. Consequently, at the intersection of RRM and NSC, there is a one-to-one mapping that links each nest to an identical constraint set. To understand why this result holds, note that, by definition, $\mu(\{x\},S)>0$ for all $x$ and $S\ni x$ when $\mu$ has a singleton representation. On the one hand, in RRM, the constraint set associated with option $x$ cannot include another alternative $y\ne x$. Otherwise, $\mu(\{x\},\xy)=0$, which contradicts  the definition of a singleton $\mu$. On the other hand, in NSC, if there exists a nest $N_i$ consisting of at least two distinct alternatives $x$ and $y$, the DM will select $\{x\}$ with zero probability at choice set $\xy$, which also contradicts the singleton $\mu$ definition.

\bigskip
\noindent \textbf{Relationship between NSC and other models.} Proposition \ref{prop: NSC-relationship} below summarizes the connections between NSC and other formulations of $\mu$. The link between NSC and RRM is omitted as it is detailed in Proposition \ref{prop: RRM-relationship}. Overall, Proposition \ref{prop: NSC-relationship} indicates that NSC is more closely aligned with other models than RRM, as the intersection of NSC and a given model always nests the intersection of RRM and that model.

\begin{prop}[NSC and other models]\label{prop: NSC-relationship} The following are equivalent: 
\vspace{-1em}
\begin{enumerate}[label=(\roman*)]
\itemsep-0.5em
    \item $\mu$ has a nest-invariant representation
    \item $\mu$ has NSC and endogenous EBA (or RCG) representations
    \item $\mu$ has an NSC representation and satisfies the probabilistic attention filter property.
\end{enumerate}
\end{prop}

Two remarks are in order. First, note that nest-invariant $\mu$ includes deterministic-with-full-choice $\mu$ and singleton $\mu$ as two particular cases. Consequently, Proposition \ref{prop: NSC-relationship} implies that the intersection of NSC and endogenous EBA (or RCG) nests the intersection of NSC and RRM (see also Proposition \ref{prop: RRM-relationship}). Second, at the intersection of NSC and endogenous EBA, the attributes in the EBA model are mutually exclusive, meaning that each option has exactly one attribute. Similarly, at the intersection of NSC and RCG, the categories in RCG are necessarily non-overlapping. These results follow directly from the partition structure of the collection of nests in the NSC model.

In summary, by combining the results in Propositions \ref{prop: logit}-\ref{prop: IC}, Propositions \ref{prop: RRM-relationship}-\ref{prop: NSC-relationship}, Remarks \ref{rem: EBA-AR-RCG}-\ref{rem: BR-relationship}, and Corollary \ref{coro: nested logit}, we establish the connections among all models of $\mu$ examined in the paper (see Figure \ref{fig: model relationship} for their relationships).


\section{Developing new models of collection choice}\label{sec:constructive}
In this section, we show that the two behavioral postulates (IIS and Relative Additivity) not only organize existing models but also help guide the construction of new, tractable models of collection choice that can accommodate violations of monotonicity and explain various context effects. Because IC is characterized by these two axioms, we use it as a benchmark for developing richer models. We present three simple examples for illustration: one relaxes IIS alone, another relaxes Relative Additivity alone, and the third relaxes both postulates.\footnote{These examples were inspired by a suggestion from an anonymous referee. We thank the referee for this helpful suggestion.}

\noindent \textbf{Preserving IIS and relaxing Relative Additivity}. Consider the following model of collection choice:
\begin{equation}\label{eq:extension_1}
  \mu^{1}(T,S)=\frac{w_{\alpha,\beta}(T)}{\sum_{\emptyset\ne T'\subseteq S}w_{\alpha,\beta}(T')}, \quad \text{where} \quad w_{\alpha,\beta}(T)=\exp\left\{\sum_{x\in T}\alpha_x+
       \sum_{\{x,y\}\subseteq T;\, x\ne y}\beta_{xy}\right\},   
\end{equation}
for some values of $\alpha_x,\beta_{xy} \in \mathbb{R}$. This model follows the Logit model of \cite{Brady-Rehbeck_2016_ECMA}, as the probability of choosing a collection is proportional to its weight. The weight function, however, takes a particular form: the weight of each collection $T$ depends on individual values of its members (measured by $\alpha$'s) and the pairwise interactions among them (measured by $\beta$'s). Parameter $\beta_{xy}$ captures the additional benefit or cost of including both $x$ and $y$ in the same collection, reflecting how well they fit together: a positive value rewards joint inclusion, while a negative value penalizes it.

If there are no pairwise interaction effects, i.e., $\beta_{xy}=0$ for every pair $(x,y)$, then $w_{\alpha,\beta}(T)=\prod_{x\in T}e^{\alpha_x}$, and this model reduces to the IC model with $\gamma(x)=\exp(\alpha_x)/(1+\exp(\alpha_x))$. For all parameters $\alpha$'s and $\beta$'s, the model satisfies IIS since the relative probabilities of two collections depend only on their weights and not on the menu. When the interaction terms are non-zero, however, the choice adjustment associated with removing an option from the menu can differ across collections, so $\mu^1$ generally violates Relative Additivity.

\smallskip
\noindent \textbf{Preserving Relative Additivity and relaxing IIS}. Consider the following model of collection choice, using the same weight function $w_{\alpha,\beta}(\cdot)$ as defined in (\ref{eq:extension_1}):
\begin{equation}\label{eq:extension_2}
 \mu^2(T,S)=
 \frac{\sum_{C:C\cap S=T}m_{\alpha,\beta}(C)}{\sum_{C:C\cap S\ne\emptyset}m_{\alpha,\beta}(C)}, \quad \text{where} \quad  m_{\alpha,\beta}(T)=\frac{w_{\alpha,\beta}(T)}{\sum_{\emptyset\ne T'\subseteq X}w_{\alpha,\beta}(T')}.
\end{equation}
This model follows Random Categorization: the DM draws a category $C$ with probability $m_{\alpha,\beta}(C)$ and chooses all its available members; the DM redraws a new category if none are available. The weight function $w_{\alpha,\beta}(\cdot)$ now describes the salience of categories of the grand set. Parameter $\alpha_x$ captures the contribution of $x$ to a category’s salience, while $\beta_{xy}$ captures the additional salience associated with grouping $x$ and $y$ together. When all interaction terms vanish, this model also reduces to the IC model.

Relative Additivity holds for the model in (\ref{eq:extension_2}), while IIS generally fails. The main difference between the two models in (\ref{eq:extension_1}) and (\ref{eq:extension_2}), and consequently between the behavioral postulates they satisfy, lies in how the interaction terms affect choice when an item is unavailable. In the former, the probability of choosing a collection containing $x$ does not depend on the interaction term $\beta_{xy}$ when $y$ is unavailable; IIS then holds because the relative weights of feasible collections are independent of the menu. In the latter, this probability may depend on $\beta_{xy}$ even when $y$ is unavailable. Consequently, irrelevant options can affect the relative choice probabilities of feasible collections through their interactions with available items, so $\mu^2$ generally violates IIS. 

This model can also explain the relative attraction effect. To illustrate, consider $X=\{x,y,z\}$, where $x$ is a target, $y$ a competitor, and $z$ a decoy similar to $x$. If $z$ interacts less strongly with $x$ than with $y$, so that $\beta_{xz}<\beta_{yz}$, adding $z$ to the menu $\{x,y\}$ increases the probability of choosing $\{x\}$ relative to that of choosing $\{y\}$.

\smallskip
\noindent \textbf{Relaxing both IIS and Relative Additivity.} Consider the following extension of the IC model:
\begin{equation}\label{eq:extension_3}
 \mu^{3}(T,S)=\frac{\prod_{x\in T}q_x(S)\prod_{y\in S\setminus T}(1-q_y(S))}
 {1-\prod_{x\in S}(1-q_x(S))}, \quad \text{where} \quad q_x(S)=\frac{\exp\{\alpha_x+\sum_{y\in S\setminus\{x\}}\beta_{xy}\}}
 {1+\exp\{\alpha_x+\sum_{y\in S\setminus\{x\}}\beta_{xy}\}}.
\end{equation}
In this extension, item-specific weights depend on the menu: each item's weight $q_x(S)$ reflects its individual score and its interactions with the other available options in the menu.\footnote{This model may arise from the menu-specific latent-scale framework of \citet{regenwetter1998random}.} This model also reduces to the IC model when all interaction terms vanish. Note that the model does not satisfy Relative Additivity or IIS in general. 

In addition to explaining the relative attraction effect, as $\mu^2$ does, this model can also explain the absolute attraction effect, where adding a decoy $z$ that is similar to $x$ but dissimilar to $y$ increases the probability of choosing the fixed collection $\{x\}$, and the similarity effect, where including $z$ in the menu reduces the probability of choosing $\{x\}$ relative to that of choosing $\{y\}$.


\section{Discussions}\label{sec:discussion}
The models studied in Sections~\ref{sec:models_and_characterization}-\ref{sec: relationship} specify a distribution over chosen collections for each menu. Their interpretation depends on whether the chosen collection is observed and whether it is the final outcome of the decision-making process or an input to a subsequent decision. This section discusses these two issues separately and then examines two related properties of $\mu$ introduced in \cite{regenwetter1998random}.

\subsection{A second stage following collection selection}\label{sec:second-stage}
In many environments, selecting the collection is the final outcome of the choice process. Examples include a voter submitting an approval ballot, a supermarket choosing an assortment to stock, and a platform selecting products to recommend. In these cases, the same decision maker makes no subsequent choice from the selected collection. While election officials may aggregate ballots and customers may buy from assortments or recommendations, these later decisions involve other decision makers and different choice situations.

Under certain interpretations, such as when $\mu$ is the distribution of consideration sets, a complete model of choice may involve a second stage in which the same decision maker chooses a single option from the selected collection. In this two-stage decision-making process, let $\rho_S(x|T)$ be the conditional probability of choosing $x$ after collection $T$ is selected from menu $S$. For each nonempty $T\subseteq S$, $\rho_S(x|T)$ forms a probability distribution supported on $T$. Let $p(x,S)$ be the probability of choosing alternative $x$ from menu $S$; it is given by
\begin{equation}\label{eq:two-stage}
p(x,S)=\sum_{\emptyset\ne T\subseteq S}\rho_S(x|T)\mu(T,S).
\end{equation}
The same $\mu$ may then produce different final choice probabilities depending on the second-stage selection rule. Two common choices of $\rho_S(\cdot|T)$ appear in the literature. Under the maximization rule, there is a strict preference ordering over all alternatives, and $\rho_S(x|T)$ equals one if $x$ is the best alternative in $T$ and zero otherwise. Under the Luce rule, each option $x$ has a positive weight $v(x)$ and its choice probability is proportional to this weight:
\[
\rho_S(x|T)=\frac{v(x)}{\sum_{y\in T}v(y)} \quad \text{for all $x\in T$}.
\]
Within the class of models with this two-stage structure, the characterization and comparison results for $\mu$ in Section~\ref{sec: relationship} help identify the sources of differences in final choice probabilities. If two models of $p$ use the same second-stage rule, different final predictions must arise from differences in collection formation. Conversely, for two models of $p$ with the same $\mu$, any differences in final choice must arise from their second-stage rules.

The Independent Choice and Random Categorization models provide a concrete illustration of the former comparison. Both models are associated with a second stage that selects the best alternative in the chosen collection according to a strict preference ordering. Holding this ordering fixed, any difference in the two models’ final choice predictions must arise from differences in $\mu$. Propositions ~\ref{prop: RCG} and ~\ref{prop: IC} then show that, when $\mu$ has full support, this difference comes down to a single property: IIS, which $\mu$ in IC necessarily satisfies and $\mu$ in RCG need not satisfy.

Sequential EBA \citep{Tversky:1972} and the Attribute Rule \citep{Gul_Natenzon_Pesendorfer_2014_ECMA} illustrate the converse comparison. \cite{Gul_Natenzon_Pesendorfer_2014_ECMA} note that the two models are distinct. Remark \ref{rem: EBA-AR} indicates that the first-stage formulations of $\mu$ in the two models are equivalent, and therefore implies that any disparity in final choice must come from distinct choice procedures after the initial collection is selected. Specifically, in an AR, conditional on the selected attribute, the DM picks an element from the chosen collection according to a Luce rule. Meanwhile, in the sequential EBA model, the DM keeps eliminating options until only one remains.

\subsection{Observability of \texorpdfstring{$\mu$}{mu}}\label{sec:observability}
In several applications and environments, the selected collection is directly observable. In recommendation settings, such as those involving vending machines, the displayed recommendations can be recorded together with the available products. \citet{kawaguchi2021designing,kawaguchi2019effectiveness} use these data in their studies. Approval ballots likewise record the selected candidates directly, and can be observed in experiments \citep{fishburn1988experiment} or through public records in real-life settings.\footnote{St. Louis’s March 4, 2025 municipal primary is one example. The data are publicly available at https://www.stlouis-mo.gov/.}

In other environments, $\mu$ may not be directly observable or easy to measure, for instance, when it represents a distribution of consideration sets. Even in these settings, the characterizations and comparisons of $\mu$ presented in the paper clarify the substantive restrictions and assumptions imposed by alternative parametric models. This helps the analyst select the parametric specification of $\mu$ most suitable for the application of interest. Additionally, in these cases, three approaches can provide information about $\mu$.

First, advances in digital tracking and eye-tracking technologies allow the analyst to collect detailed records of the decision-making process by monitoring individuals' clicks, information searches, viewing patterns, the time spent examining each piece of information, and eye movements \citep{reutskaja2011search,lahey2016power,chaturapruek2021studying}. In such environments, these records can help reconstruct selected collections and their distribution under appropriate assumptions.

Second, the analyst can ask decision makers to directly report the collections they select from given menus through surveys, web-based interviews, and laboratory experiments \citep{brown1992consideration,hauser2010disjunctions,EllisOzbayFilizCC}. For instance, \citet{EllisOzbayFilizCC} design a lab experiment in which participants report which alternatives they would consider from different menus. They repeat each menu to estimate the frequency of each reported collection.

Finally, in situations in which the selection of a collection is followed by an observed final choice from it, the analyst can estimate $\mu$ using the final choice data under suitable identifying assumptions \citep{chiang1998markov,hwang2010empirical,abaluck2021consumers}. For instance, \citet{hwang2010empirical} reconstruct supermarket assortments from store-level scanner sales data, while \cite{chiang1998markov} estimate the distribution of consideration sets using household scanner panel data on choices among four ketchup brands.


\subsection{Related properties}\label{sec:substructure}
\citet[Definition~11]{regenwetter1998random} examine whether projecting choice probabilities onto smaller menus preserves their representability within a class of models. Let $\mathcal{M}$ denote a generic class of models. Let $\mu(\cdot,X)$ denote the observed choice probabilities on the grand set $X$, and assume that every alternative belongs to at least one collection $T$ with $\mu(T,X)>0$. Suppose that these choice probabilities admit a representation in $\mathcal{M}$. For each nonempty menu $S\subseteq X$, choice probabilities in $S$ may be unobserved; we construct them by projecting the choice probabilities in the grand set onto $S$. In our domain of nonempty collection choice, the resulting probabilities are
\[
\widehat{\mu}(T,S)=\frac{\sum_{T'\cap S=T}\mu(T',X)}
{\sum_{T'\cap S\ne\emptyset}\mu(T',X)},
\qquad \emptyset\ne T\subseteq S.
\]
This $\widehat{\mu}$ is well defined for every $(T,S)$ such that $\emptyset\ne T\subseteq S$, as the denominators are strictly positive, which follows from the assumption that every alternative appears in some collection selected with positive probability from the grand set. The class $\mathcal{M}$ is then said to be stable under substructures if, for every given $\mu(\cdot,X)$ that has a representation in $\mathcal{M}$, $\widehat{\mu}$ also has a representation in $\mathcal{M}$. This means that there exists a fixed, single representation in $\mathcal{M}$ that generates $\widehat{\mu}$ across all menus. Stability under substructures thus concerns whether the projected data remain explainable by the same class of models. Note that stability under substructures is a property of a class of models, while the two main behavioral postulates in the paper (IIS and Relative Additivity) are properties of an individual choice system.

Another property of $\mu$ studied in \cite{regenwetter1998random} is the
\textit{context-invariant general constant ratio condition}.
When the chosen collection is nonempty, we say
that $\mu$ satisfies this condition if, for every pair of menus $S,S'$, every $x\in S\cap S'$, and all
nonempty $T\subseteq S\setminus\{x\}$ and $T'\subseteq S'\setminus\{x\}$,
\begin{equation}\label{eq:gcrc-nonempty}
\mu(T\cup\{x\},S)\mu(T',S') = \mu(T'\cup\{x\},S')\mu(T,S).
\end{equation}
Whenever $\mu(T,S)$ and $\mu(T',S')$ are positive, this condition
is equivalent to $\mu(T\cup\{x\},S)/\mu(T,S)= \mu(T'\cup\{x\},S')/\mu(T',S')$. Thus, the probability of selecting
a collection containing $x$, relative to the probability of selecting
the same collection without $x$, is independent
of both the other alternatives in that collection and the menu. Unlike IIS, which compares the probabilities of two fixed collections across
menus, this condition compares the odds of adding a given alternative to possibly different collections, both within and across menus. \cite{regenwetter1998random} use this condition to provide an alternative characterization of IC.\footnote{More precisely, their characterization is stated on a domain allowing empty collections, but it extends to our
full-support domain in which the chosen collection must be nonempty.}

Table \ref{tab:properties} reports whether the models studied in the paper are stable under substructures and satisfy the context-invariant general constant ratio condition. It suggests a connection between stability under substructures and Relative Additivity: RCG, IC, and static EBA all satisfy Relative Additivity and are stable under substructures, whereas Logit does not satisfy Relative Additivity and is not stable under substructures. RRM and NSC are exceptions to this general pattern: NSC is stable under substructures even though its choice rules need not satisfy Relative Additivity, while RRM satisfies a generalized form of Relative Additivity but is not stable under substructures.

\begin{table}[ht!]
\centering
\caption{Models and behavioral postulates}
\label{tab:properties}
\footnotesize
\setlength{\tabcolsep}{3pt}
\renewcommand{\arraystretch}{1.25}

\begin{tabular*}{\textwidth}{
    @{\extracolsep{\fill}}
    >{\raggedright\arraybackslash}m{0.42\textwidth}
    ccc|c|ccc@{}
}
\hline
\hline
\textbf{Properties} & \multicolumn{7}{c}{\textbf{Models}} \\
\cline{2-8}
& \multicolumn{3}{c|}{List A} &  \multicolumn{1}{c|}{List B} & \multicolumn{3}{c}{List C} \\
\cline{2-8}
& Logit & \shortstack{Nested\\logit} & NSC & IC
& \shortstack{Static\\EBA} & RCG & RRM \\
\hline
Independence of Irrelevant Sets
& $\checkmark$ & $\checkmark$ & $\checkmark$ & $\checkmark$
& $\times$ & $\times$ & $\times$ \\

Path-Independence of Irrelevant Sets
& $\checkmark$ & $\checkmark$ & $\checkmark$ & $\checkmark$
& $\times$ & $\times$ & $\times$ \\

\hline
Relative Additivity
& $\times$ & $\times$ & $\times$ & $\checkmark$
& $\checkmark$ & $\checkmark$ & $\times$ \\

Relative Additivity-1
& $\times$ & $\checkmark$ & $\checkmark$ & $\checkmark$
& $\checkmark$ & $\checkmark$ & $\checkmark$ \\

\hline
Stability under substructures
& $\times$ & $\times$ & $\checkmark$ & $\checkmark$
& $\checkmark$ & $\checkmark$ & $\times$ \\

Context-invariant general constant ratio
& $\times$ & $\times$ & $\times$ & $\checkmark$
& $\times$ & $\times$ & $\times$ \\
\hline
\hline
\end{tabular*}

\medskip
\parbox{\textwidth}{\footnotesize
\textit{Notes:} $\checkmark$ indicates that every member of the
model class satisfies the stated property; $\times$ indicates
failure in general. IIS and PIIS have identical entries for these
model classes but are not equivalent in general. Static EBA uses exogenously given attributes. For the nested logit model of collection choice, see
Appendix~\ref{appendix:nested_logit}.}
\end{table}

\newpage
\section{Related literature} \label{sec: Literature Review}
Our paper primarily contributes to two strands of the literature. The first studies random consideration set formation, pioneered by \cite{Manzini-Mariotti_2014_ECMA} and \cite{cattaneo2020random}, among others. Our work contributes to this literature by demonstrating the link between various parametric models of random consideration set formation and by offering a practical tool, based on behavioral postulates, to test and potentially falsify these models. 

The work closest to ours in this branch of literature is \cite{kovach2023reference}. Their paper assumes that a reference point affects the formation of $\mu$ and interprets $\mu$ as a distribution of consideration sets. Their focus differs fundamentally from ours: we investigate choices that involve selecting multiple options, whereas their study analyzes how reference-dependent attention affects the choice probabilities of individual options. Moreover, their paper does not provide a behavioral characterization of any model of $\mu$ examined in our study. 

Another broadly related paper is \cite{manzini2024model}, which proposes a model of approval in which decision makers approve a collection of options for later use from a ranked list of items. \cite{manzini2024model} introduce a new model rather than providing a systematic examination of well-known models of $\mu$, as we do. Additionally, items in our setting are not necessarily ordered.

As the $\mu$ object studied in the paper is a stochastic choice correspondence, our work contributes to a second strand of the literature focusing on axiomatizing choice correspondences. Most papers in this literature focus on deterministic environments.\footnote{See, for instance, \cite{AizermanAleskerov1995, masatlioglu2005rational, salant2008f}, among others.} Our paper belongs to the small set of studies axiomatizing stochastic choice correspondences (see also \cite{barbera1986falmagne} and \cite{regenwetter1998random}). Additionally, the decision maker in our framework deliberately chooses multiple alternatives each time she faces a choice problem. Meanwhile, the literature on deterministic choice correspondences typically assumes that the DM selects a single option, and a choice correspondence arises from factors such as indifference, incompleteness, multiple preferences \citep{rubinstein2012lecture, balakrishnan2022inference}, or frames \citep{salant2008f}.

The work closest to ours in this branch of the literature is \citet{regenwetter1998random}. This paper and our study are complementary but differ in three key aspects: (i) their focus, (ii) their approach, and (iii) their results.

First, regarding the focus, \citet{regenwetter1998random} primarily investigate the class of random utility threshold models where the decision maker chooses all available options whose utilities meet or exceed a possibly random threshold. Their Theorem 6 (p. 181) establishes that all random utility threshold models are stable under substructures. Our paper focuses on mostly parametric models of collection choice that are used in recent economic literature, with choice rules arising from collection weights, observed attributes, feasible references, and partitions of alternatives. As shown in Table \ref{tab:properties}, several models studied in our paper are not stable under substructures, further indicating a difference in focus between the two papers.

Second, regarding the approach, \citet{regenwetter1998random} start with a general random utility threshold framework and study special cases and related representations. These include (i) the latent-scale model in which utilities of options are independent and the threshold is fixed; (ii) the context-invariant threshold model in which the joint distribution of utilities and the threshold remains unchanged across menus; and (iii) the size-independent model in which the decision maker first decides how many options to choose and then selects that many options with the highest utilities. Their analysis compares models formulated within a common utility-based framework and examines their common properties and relationships. Our approach begins with seemingly unrelated models of collection choice motivated by distinct economic mechanisms and expressed in different functional forms. We not only provide behavioral characterizations but also identify shared postulates and use them to organize these models through a common axiomatic framework. Through these behavioral postulates, we establish relationships among models that are not transparent from the models’ original formulations.

Finally, regarding the results, the most relevant findings for our paper in \citet{regenwetter1998random} are the equivalence between independent random utility constant-threshold models and latent-scale models (Theorem 2, p. 178), and the characterization of the context-invariant latent-scale model by the context-invariant general constant ratio condition (Theorem 3, p. 178). When $\mu$ has full support on the nonempty-collection domain, the context-invariant latent-scale model is indeed the IC model. Their results thus connect a parametric representation to a non-parametric representation. In their characterization for IC, they treat the context-invariant general constant ratio condition as a standalone characterization, without exploring its connection to other properties studied in their paper. We provide an alternative characterization of IC through the joint imposition of IIS and Relative Additivity, thereby decomposing their context-invariant general constant ratio condition. The decomposition is useful for connecting this condition to other more standard properties, and for constructing new models of collection choice, as illustrated in Section \ref{sec:constructive}.


\section{Conclusion} \label{sec: Conclusion}
We study choice situations that naturally involve selecting multiple options. We systematically investigate various models of stochastic collection choice in the literature by providing their behavioral foundations and identifying their relationships. Theoretically, our results indicate that popular models, though based on distinct underlying narratives, share similar behavioral foundations. These theoretical results provide a behavioral taxonomy of stochastic collection-choice models, help the analyst better understand the implications of the functional form assumptions, and distinguish between different models of $\mu$. Empirically, our behavioral postulates facilitate testing which model classes are consistent with decision makers' observed collection choices. More broadly, our results provide the analyst with a common framework for understanding, comparing, and evaluating models of collection choice, with applications ranging from consumer consideration sets and recommendation systems to portfolio selection and approval voting.

\bigskip
\bigskip
\bigskip
\bigskip

\bibliographystyle{ecta}
\bibliography{references1.bib}

\clearpage
\normalsize
\begin{appendices}

\section{Additional results and models}\label{appendix:additional_results_and_models}

\subsection{Characterization of RRM}\label{appendix:RRM_characterization}
This appendix provides a characterization of $\mu_{\texttt{RRM}}$. We adopt a \textit{revealed constraint approach.} We first assume that $\mu$ has an RRM representation and identify the constraint sets from $\mu$. We subsequently utilize the identified constraint sets in behavioral postulates. Identifying the constraint sets uses choice probabilities from binary choice sets. Table \ref{tab: identification-RRM}  presents the probabilities of choosing different collections at choice set $\xy$ under four possible scenarios regarding the constraint sets. 
\begin{table}[ht!]
    \centering
    \begin{tabular}{c|c|c|c}
\toprule
         Scenarios &  $\mu_{\texttt{RRM}}(\xy,\xy)$ & $\mu_{\texttt{RRM}}(\{x\},\xy)$ & $\mu_{\texttt{RRM}}(\{y\},\xy)$ \\
    \hline
         $x\in \mathcal{Q}(y)$ and $y\in \mathcal{Q}(x)$ & 1 & 0 & 0   \\
        $x\not\in \mathcal{Q}(y)$ and $y\not\in \mathcal{Q}(x)$ & 0 & $\frac{s_x}{s_x+s_y}$ & $\frac{s_y}{s_x+s_y}$   \\
    $x\not\in \mathcal{Q}(y)$ and $y\in \mathcal{Q}(x)$ & $\frac{s_x}{s_x+s_y}$ & $0$ & $\frac{s_y}{s_x+s_y}$   \\
        $x\in \mathcal{Q}(y)$ and $y\not\in \mathcal{Q}(x)$ & $\frac{s_y}{s_x+s_y}$ & $\frac{s_x}{s_x+s_y}$ & $0$   \\
\bottomrule
\end{tabular}
    \caption{Probabilities of choosing different collections in choice set $\xy$ in RRM}
    \label{tab: identification-RRM}
\end{table}

Suppose $\mu$ has an RRM representation. From Table \ref{tab: identification-RRM}, it is immediate that $y$ belongs to the constraint set associated with $x$ if and only if the DM selects the singleton set $\{x\}$ from binary choice set $\xy$ with zero probability. To understand why this property holds under the RRM, regarding the if part, note that if $y\not \in \mathcal{Q}(x)$, then the DM chooses $\{x\}$ from choice set $\xy$ when the reference point is $x$. This implies that $\mu(\{x\},\xy)>0$, which contradicts the assumption that $\mu(\{x\},\xy)=0$. For the only-if part, by definition of the RRM, the DM chooses $\mathcal{Q}(x)\cap \xy$ or $\mathcal{Q}(y)\cap \xy$ when $x$ or $y$ serves as the reference point, respectively. The assumption that $y\in \mathcal{Q}(x)$ implies $y\in \mathcal{Q}(x)\cap \xy$ and the fact that $y$ is included in its constraint set implies $y\in \mathcal{Q}(y)\cap \xy$. Hence, regardless of which option serves as the reference point, the DM never chooses $\{x\}$. It follows that the choice probability of $\{x\}$ in choice set $\xy$ must be equal to $0$. This observation motivates the definition of the revealed constraint function $\mathcal{Q}^R$ in the main text.

In addition to Axiom \ref{ax: RRM1}, the characterization of RRM requires the following three axioms. The first axiom (Axiom \ref{ax: RRM-distinct}) follows from the assumption that the constraint sets associated with two different alternatives are distinct. 

\begin{ax}[Distinct Constraint Sets]\label{ax: RRM-distinct} $\mathcal{Q}^R(x)\ne \mathcal{Q}^R(y)$ when $x\ne y$.
\end{ax}

The second axiom (Axiom \ref{ax: RRM-Positivity}) states that the DM chooses a collection $T$ with a positive probability if and only if there exists a reference point $x$ such that $T$ is identical to the collection of all feasible alternatives that belong to the constraint set associated with $x$. Axiom \ref{ax: RRM-Positivity} is a direct consequence of the definition of RRM. Compared to RCG and EBA models, Axiom \ref{ax: RRM-Positivity} is similar to Axiom \ref{ax: RCG-Positivity} and Axiom \ref{ax: EBA-posi}.

\begin{ax}[Positivity-3] \label{ax: RRM-Positivity} For all non-empty $T\subseteq S$, $\mu(T,S)>0$ if and only if there exists some $x\in S$ such that $T= \mathcal{Q}^R(x)\cap S$. 
\end{ax}

The next behavioral postulate imposes a restriction on $\mu$ when $T=\mathcal{Q}^R(x) \cap (S\setminus x)$, a case not covered by Axiom \ref{ax: RRM1}. It is also a variant of Relative Additivity.

\begin{ax}[Relative Additivity-2]\label{ax: RRM2} Suppose $x\in S$ and $T=\mathcal{Q}^R(x) \cap (S\setminus x)$ is non-empty. Suppose $\sum_{y\in S} \mu(\mathcal{Q}^R(y),X)>0$. For all non-empty $T'\subseteq S\setminus x$ such that $T'\ne T$
\footnotesize
\begin{equation}\label{eq: RRM-ax}
   \mu(T,S\setminus x)[\mu(T',S)+\mu(T'\cup x, S)]=\mu(T',S\setminus x) \Big[\mu(T,S)+\mu(T\cup x, S)\underbrace{-\frac{\mu(\mathcal{Q}^R(x),X)}{\sum_{y\in S} \mu(\mathcal{Q}^R(y),X)}}_{\text{adjustment term}}\Big]. 
\end{equation}
\end{ax}
\normalsize

Axiom \ref{ax: RRM2} primarily differs from the original Relative Additivity and Axiom \ref{ax: RRM1} through the inclusion of an additional term, $-\frac{\mu(\mathcal{Q}^R(x),X)}{\sum_{y\in S} \mu(\mathcal{Q}^R(y),X)}$, on the right-hand side of equation (\ref{eq: RRM-ax}). This adjustment term, in absolute value, represents the probability of $x$ being the reference point under $S$, given that the salience of $x$ is identified as $s_x=\mu(\mathcal{Q}^R(x), X)$. Including this adjustment term, as in Axiom \ref{ax: RRM2}, neutralizes the influence of $x$ on the choice probability of $T\cup x$ when serving as the reference point at $S$.

We are now ready to state a characterization of RRM.

\begin{prop}[Characterization of RRM]\label{prop: RRM} $\mu$ has an RRM representation if and only if it satisfies Axiom \ref{ax: RRM1} and Axioms \ref{ax: RRM-distinct}-\ref{ax: RRM2}. 
\end{prop}

Proposition \ref{prop: RRM} provides the analyst with a practical means to test whether $\mu$ has an RRM representation by validating four simple conditions. The proof of Proposition \ref{prop: RRM} proceeds as follows. We first define the constraint function $\mathcal{Q}$ as the \textit{revealed constraint function $\mathcal{Q}^R$} and the salience of option $x$ as $s_x=\mu(\mathcal{Q}^R(x), X)$. Axioms \ref{ax: RRM-distinct} and \ref{ax: RRM-Positivity} guarantee that $\{\mathcal{Q}(x)\}_x$ and $\{s_x\}_x$ are well defined. We subsequently show that $\mu$ has an RRM representation given the defined constraint sets and salience parameters. The proof uses induction based on the number of alternatives in the choice set. The proof of Proposition \ref{prop: RRM} also provides a useful identification result, demonstrating that the constraint function is uniquely identified, and the salience parameters are unique up to uniform scaling.


\subsection{Characterization of NSC}\label{appendix:NSC_characterization}
Besides PIIS, characterizing NSC requires two additional behavioral postulates (Axioms \ref{ax: NSC-Partition}-\ref{ax: NSC-Positivity}). Axiom \ref{ax: NSC-Partition} follows from the partition structure of the nests in the model's definition. Given observed data on $\mu$, let $N^R_1,N^R_2,\dots,N^R_{q}$ be all pairwise distinct collections selected with positive probability in the grand set ($q\ge 1$). These collections are the \textit{revealed nests}, i.e., the underlying nests if $\mu$ has an NSC representation. Axiom \ref{ax: NSC-Partition} states that these \textit{revealed nests} form a partition of the grand set.

\begin{ax}[Partition] \label{ax: NSC-Partition} $N_i^R\cap N_j^R=\emptyset$ for all $i,j$ such that $i\ne j$. Furthermore, $\bigcup_{i=1}^{q} N_i^R=X$.
\end{ax}

Axiom \ref{ax: NSC-Positivity}, presented below, is similar to Positivity-1 (Axiom \ref{ax: RCG-Positivity}) in RCG, Positivity-2 (Axiom \ref{ax: EBA-posi}) in EBA, or Positivity-3 (Axiom \ref{ax: RRM-Positivity}) in RRM. Axiom \ref{ax: NSC-Positivity} states that the DM chooses a collection with a strictly positive probability if and only if it corresponds to the set of feasible options within some \textit{revealed nest}. This behavioral postulate is also a consequence of the model's definition.

\begin{ax}[Positivity-4] \label{ax: NSC-Positivity} For all non-empty $T\subseteq S$, $\mu(T,S)>0$ if and only if there exists some $i$ such that $T=N_i^R\cap S$.
\end{ax}

We are now ready to state a characterization of NSC. Proposition \ref{prop: NSC} establishes that NSC is characterized by PIIS, Partition, and Positivity-4.

\begin{prop}[Characterization of NSC] \label{prop: NSC} $\mu$ has an NSC representation if and only if it satisfies PIIS, Partition, and Positivity-4. 
\end{prop}

Proposition \ref{prop: NSC} implies that determining whether $\mu$ has an NSC representation is equivalent to testing whether it satisfies PIIS, Partition, and Positivity-4. The proof of sufficiency in Proposition \ref{prop: NSC} proceeds as follows. Define the nests as \textit{the revealed nests.} Obtaining an NSC representation is trivial when there is only one revealed nest. When at least two distinct revealed nests exist, we construct the weighting function $\sigma$ using a cross-nest strategy. Fix two arbitrary options in two different revealed nests: $x_1\in N^R_1$ and $x_2\in N^R_2$. Assign $\sigma(\{x_1\})=1$ and let $\sigma(\{x_2\})=\frac{\mu(\{x_2\},\{x_1,x_2\})}{\mu(\{x_1\},\{x_1,x_2\})}$. When $T$ and $\{x_1\}$ are in different revealed nests, define $\sigma(T)$ through $\sigma(\{x_1\})$ by setting $\sigma(T)=\frac{\mu(T,T\cup x_1)}{\mu(\{x_1\},T\cup x_1)}\sigma(\{x_1\})$. When $T$ and $\{x_1\}$ are in the same revealed nest, we use the fact that $T$ and $\{x_2\}$ are in two different nests and define $\sigma(T)$ through $\sigma(\{x_2\})$ by setting $\sigma(T)=\frac{\mu(T,T\cup x_2)}{\mu(\{x_2\},T\cup x_2)}\sigma(\{x_2\})$. Using PIIS, Partition, and Positivity-4, we prove that the $\sigma$ function is well defined and that $\mu$ has an NSC representation given $\sigma$ and the revealed nests. The proof of Proposition \ref{prop: NSC} also indicates that if $\mu$ has an NSC representation, then the nests can be uniquely identified (up to relabeling) and the weighting function is unique up to uniform scaling on within-nest collections (when $q\ge 2$).

\subsection{Logit, RCG, and IC with empty chosen collection}\label{sec: empty choices}
In the main body of the paper, we investigated Logit, RCG, and IC under the assumption that the chosen collection is non-empty. In this section, we relax this assumption and provide characterizations for the three models when the chosen collection is allowed to be empty. We add a superscript $^o$ to the names of these models to differentiate them from the models studied in the main body of the paper. Table \ref{tab: empty choice} provides the functional forms of the three models when the chosen collection can be empty.
\begin{table}[ht!]
    \centering
    \begin{tabular}{l|l}
    \hline
    \hline
    \text{Models} & \text{Functional forms when the chosen collection can be empty} \\
    \hline
         \hline
        Logit$^o$ &  $\mu^o_{\texttt{LG}}(T,S)=\frac{\pi(T)}{\sum_{T':T'\subseteq S}\pi(T')} \quad \text{ for all }\, T\subseteq S$ \\[9pt]
         RCG$^o$ &  $\mu^o_{\texttt{RCG}}(T,S)=\sum_{C} m(C) \mathbbm{1}(T=S \cap C) \quad \text{ for all }\, T\subseteq S$ \\[9pt]
         \hline
         IC$^o$ & $\mu^o_{\texttt{IC}}(T,S)=\prod_{x\in T}\gamma(x) \prod_{y\in S\setminus T}(1-\gamma(y)) \quad \text{ for all }\, T\subseteq S$ \\[9pt]
         \hline
         \hline
    \end{tabular}
    \caption{Logit, RCG, and IC with empty chosen collections}
    \label{tab: empty choice}
\end{table}

\noindent \textbf{Characterization of RCG$^o$}. Under the new assumption that the chosen collection can be empty, RCG$^o$ is characterized by a single behavioral postulate, Additivity (Axiom \ref{ax: RCG*}). Additivity states that removing an alternative $x$ from the choice set weakly increases the choice probability of any collection $T$ that excludes $x$, with the increase being precisely equal to the likelihood that $T \cup {x}$ is the chosen collection in the original choice set. 

\begin{ax}[Additivity]\label{ax: RCG*} For all $x,T,S$ such that $x\in S$ and $T\subseteq S\setminus x \ne \emptyset$
\[
\mu(T,S\setminus x)-\mu(T,S)=\mu(T\cup x, S).
\]
\end{ax}
Note that Additivity implies Relative Additivity (Axiom \ref{ax: RCG}) and is therefore more restrictive. To elaborate, Additivity implies $\mu(T,S\setminus x)=\mu(T,S)+\mu(T\cup x, S)$ and $\mu(T',S)+\mu(T'\cup x, S)=\mu(T',S\setminus x)$. Multiplying these two equations yields Relative Additivity. Additionally, one can show that Additivity is equivalent to the Consistency under marginalization condition in \cite{regenwetter1998random}.

Proposition \ref{prop: RCG*} establishes that Additivity characterizes the RCG$^o$ model. 

\begin{prop}[Characterization of RCG$^o$] \label{prop: RCG*} $\mu$ has an RCG$^o$ representation if and only if it satisfies Additivity. 
\end{prop}

\noindent \textbf{Characterization of IC$^o$}. With regard to the IC$^o$ model, Proposition \ref{prop: IC*} establishes that a full-support $\mu$ has an IC$^o$ representation if and only if it satisfies IIS$^o$ and Additivity. Here, IIS$^o$ is the variant of IIS (Axiom \ref{ax: IIS}) that allows the chosen collection to be empty.\footnote{Formally, IIS$^o$ states that $\mu(T,S)/\mu(T',S)=\mu(T,S')/\mu(T',S')$ for all $T,T' \subseteq S\cap S'$ whenever the two ratios are well defined. IIS$^o$ fully characterizes the Logit$^o$ model when $\mu$ has full support. Because the characterization of Logit$^{o}$ deviates only slightly from that of Logit, we omit it here for brevity.}

\begin{prop}[Characterization of IC$^o$] \label{prop: IC*} A full-support $\mu$ has an IC$^o$ representation if and only if it satisfies IIS$^o$ and Additivity. 
\end{prop}

As in the environment where the chosen collection must be non-empty, Proposition \ref{prop: IC*} implies that the IC$^o$ model remains equivalent to the intersection of the Logit$^o$ and RCG$^o$ models. This result is stated in Corollary \ref{coro: IC*}.

\begin{coro}\label{coro: IC*} Suppose $\mu$ has full support. The following are equivalent: 
\vspace{-0.3cm}
\begin{enumerate}[label=(\roman*)]
\itemsep0em
\item $\mu$ has an IC$^o$ representation
\item $\mu$ has Logit$^o$ and RCG$^o$ representations.
\end{enumerate}
\end{coro}


\subsection{Attribute rule}\label{sec: Attribute Rule}
In a framework related to the original EBA model, \cite{Gul_Natenzon_Pesendorfer_2014_ECMA} introduce an Attribute Rule model to study choice situations where the decision maker selects a single option. In an Attribute Rule (AR), the DM first picks an attribute according to a logit formula and subsequently chooses an alternative possessing that attribute according to another logit formula. Let $B(S)$ represent the set of all attributes that are possessed by at least one element in $S$. The probability of choosing an option $x$ from $S\ni x$ is denoted as $p_{\texttt{AR}}(x,S)$ and is given by 
\[
p_{\texttt{AR}}(x,S)=\sum_{i\in B(S)}\frac{\theta_i}{\sum_{j\in B(S)}\theta_j} \frac{\eta^i_x}{\sum_{y\in S}\eta^i_y} \quad \text{ for all } x\in S.
\]
In the formulation of $p_{\texttt{AR}}(x,S)$, function $\theta$ governs the choice of attributes and maps attributes to $\mathbb{R}_{++}$. Meanwhile, function $\eta$ governs the choice of options and maps a pair of an attribute and an option possessing that attribute to the natural numbers, with $\eta^i_x>0$ if option $x$ has attribute $i$ and $\eta^i_x=0$ otherwise.

Let $B_i$ denote the collection of options possessing attribute $i$. As $\eta^i_x=0$ if $x$ does not have attribute $i$, it is the case that $\sum_{y\in S}\eta^i_y =  \sum_{y\in (B_i\cap S)}\eta^i_y$. Equivalently,
\[
p_{\texttt{AR}}(x,S)=\sum_{i\in B(S)}\frac{\theta_i}{\sum_{j\in B(S)}\theta_j} \frac{\eta^i_x}{\sum_{y\in (B_i\cap S)}\eta^i_y} \quad \text{ for all } x\in S.
\]
This reformulation motivates an alternative interpretation of an Attribute Rule as follows. The decision maker first draws an attribute $i$ with probability $\frac{\theta_i}{\sum_{j\in B(S)}\theta_j}$. The DM then pays attention only to feasible options possessing that attribute, which is the set $T=B_i\cap S$. Finally, the DM picks an option $x$ from $T$ with probability proportional to its value on attribute $i$. Based on this alternative interpretation of the AR model, we can reformulate it as a two-stage decision-making process. In the first stage the DM chooses a collection $T$ from the choice set $S$; in the second stage she selects an option from $T$ as the final choice. 

Mathematically, let $I_{T,S} = \{i:i \in B(S) \text{ such that } T=B_i \cap S\}$ denote the set of attributes that induce collection $T$ to be chosen in the first stage. The probability of choosing $T$ from choice set $S$ equals the total weight of all attributes $i$ in $I_{T,S}$, divided by the total weight of all feasible attributes:
\begin{equation}\label{eq: AR1}
\mu_{\texttt{AR}}(T,S)=\sum_{i\in I_{T,S}}\frac{\theta_i}{\sum_{j\in B(S)} \theta_j} \quad \text{ for all non-empty } T\subseteq S. 
\end{equation}

In the second stage, the DM selects $x$ from $T$ with probability $\rho_S(x|T)$. When $I_{T,S}\ne \emptyset$, it is defined as
\[
\rho_S(x|T)=\mathbb{E}_{i\in I_{T,S}} [\Pr(x|T,i)] = \sum_{i \in I_{T,S}}[\Pr(i|T,S)\cdot \Pr(x|T,i)],
\]
with $\rho_S(x|T)=0$ whenever $x\not\in T$.\footnote{When $I_{T,S}=\emptyset$, let $\rho_S(\cdot|T)$ be an arbitrary probability distribution supported on $T$.} Here, $\mathbb{E}_{i\in I_{T,S}}$ denotes expectation over $i \in I_{T,S}$. The term $\Pr(x|T,i)$ denotes the probability that an option $x$ is chosen from $T$ based on its value on attribute $i$ and is given by $\Pr(x|T,i)= \frac{\eta^i_x}{\sum_{y \in T}\eta^i_y}$. Meanwhile, $\Pr(i|T,S)$ represents the probability that attribute $i$ induced collection $T$ to be chosen in the first stage. This probability is given by $\Pr(i|T,S)=\frac{\theta_i}{\sum_{j \in I_{T,S}} \theta_j}$ whenever $I_{T,S}\ne \emptyset$. Note that $\rho_S(x|T)$ generally depends on $S$ as the set $I_{T,S}$ depends on $S$; we therefore make the subscript $S$ explicit in $\rho_S(x|T)$.

It is straightforward to verify that, for each non-empty $S\subseteq X$ and each non-empty $T\subseteq S$, $\rho_S(.|T)$ constitutes a probability distribution over $T$, and that
\[
p_{\texttt{AR}}(x,S)= \sum_{T:\,T\subseteq S, T\ne \emptyset} \mu_{\texttt{AR}}(T,S)\cdot \rho_S(x|T) \quad \text{ for all } x\in S, \text{ for all } S\subseteq X.
\]
We say that $\mu$ has an AR representation if it has the form given in equation (\ref{eq: AR1}), i.e., if it corresponds to the first stage of the choice procedure in an AR after reformulation. Note that the functional forms of $\mu_{\texttt{EBA}}$ in equation (\ref{eq: EBA1}) and $\mu_{\texttt{AR}}$ in equation (\ref{eq: AR1}) are identical. To see this, define $\omega_i=\theta_i$ and $A_i=B_i$ for all aspects $i$. This equates the two numerators in $\mu_{\texttt{EBA}}$ and $\mu_{\texttt{AR}}$. Regarding the two denominators, note that $A_j\cap S \ne \emptyset$ if and only if there is at least one option in $S$ possessing aspect $j$. This is equivalent to saying $j\in B(S)$. Hence, the formulations of $\mu$ in static EBA and AR are identical. This result is stated in Remark \ref{rem: EBA-AR}.

\begin{rem}\label{rem: EBA-AR} $\mu$ has a static EBA representation if and only if it has an AR representation. 
\end{rem}


\subsection{Nested logit collection choice}\label{appendix:nested_logit}

In this section, we introduce the formulation of $\mu$ in the nested logit model and present its connection to other models we investigated. Nested logit is a special case of the NSC model. Given a choice set $S$, the probability of choosing a non-empty $T\subseteq S$ is defined as
\[
\mu_{\texttt{NL}}(T,S)=\sum_{i=1}^q\frac{\sigma(N_i\cap S)}{\sum_{j=1}^q \sigma(N_j\cap S)}\mathbbm{1}(T=N_i\cap S) \quad \text{ for all non-empty } T\subseteq S,
\]
where $(N_1,\dots,N_q)$ is a partition of the grand set and $\sigma$ is the weighting function. Nested logit differs from NSC in that there exist $\eta_1,\dots,\eta_q \in \mathbb{R}_{++}$ and a function $v\colon X\rightarrow \mathbb{R}_{++}$ such that $\sigma(N_i\cap S)=\big(\sum_{x\in (N_i\cap S)} v(x)\big)^{\eta_i}$ for every nest $N_i$ and non-empty $S\subseteq X$. The function $v$ represents the DM's preference over alternatives. Meanwhile, the scalar $\eta_i$ captures the degree of substitutability or complementarity among elements in nest $N_i$. As the value of each collection, $\sigma(T)$, is increasing in the total utilities of its elements, $\sum_{x\in T} v(x)$, the probability of choosing a collection depends on the DM's underlying preference over alternatives. The stronger the preference for elements within the collection, the higher the probability that the collection is selected. 

We assume that there are at least two nests ($q\ge2$) in the nested logit collection choice model. With a single nest ($q=1$), NSC and nested logit are observationally equivalent: in both models, the decision maker selects the entire available menu with probability one.

Corollary \ref{coro: nested logit} describes the relationship between $\mu$ in nested logit and the other models studied in our paper. It shows that nested logit is highly distinct from those models. The corollary follows from Propositions \ref{prop: RRM-relationship} and \ref{prop: NSC-relationship}.

\begin{coro}\label{coro: nested logit} The following are equivalent:
\begin{enumerate}[label=(\roman*)]
\itemsep-0.3em
    \item $\mu$ has a singleton representation
    \item $\mu$ has nested logit and RRM representations
    \item $\mu$ has nested logit and endogenous EBA (or RCG) representations.
\end{enumerate}
\end{coro}

\newpage
\section{Omitted proofs}\label{appendix: proofs}
\normalsize

\noindent \textbf{Proof of Proposition \ref{prop: logit}:} The necessity is straightforward. We show the sufficiency. Define $\pi(T)=\mu(T,X)$ for all non-empty $T\subseteq X$. As $\mu$ has full support, $\pi(T)>0$ for all non-empty $T\subseteq X$. By definition, $\frac{\pi (T)}{\pi (T')}=\frac{\mu(T,X)}{\mu(T',X)}$ for all non-empty $T,T'\subseteq X$. It immediately follows that $\frac{\pi(T)}{\sx \pi(T')}=\frac{\mu(T,X)}{\sx \mu(T',X)}=\mu(T,X)$ for all non-empty $T \subseteq X$, where the second equation uses the fact that $\sx \mu(T',X)=1$. Now, consider an arbitrary choice set $S\subseteq X$ such that $S\ne \emptyset$. We have
\[
\frac{\mu(T,S)}{\mu(T',S)}=\frac{\mu(T,X)}{\mu(T',X)}=\frac{\pi (T)}{\pi (T')} \quad \text{ for all non-empty } T,T'\subseteq S.
\]
The first equation holds because of the IIS axiom. The second equation holds because of the definition of $\pi$. It follows from the equation above that $\mu(T,S)=\frac{\mu(T,S)}{\sz \mu(T',S)}=\frac{\pi(T)}{\sz \pi(T')}$. 
Hence, $\mu$ has a Logit representation. This completes our proof of Proposition \ref{prop: logit}. $\blacksquare$

\bigskip

\noindent \textbf{Proof of Proposition \ref{prop: RCG}:} We first state and prove the following Lemma.

\begin{lemma}\label{lema: MA} Suppose $\mu$ satisfies Positivity-1 and Relative Additivity. Then for all $S\ni x$, $|S|\ge 2$, and all non-empty $T\subseteq S\setminus x$
\begin{itemize}
    \item [(i)] $\mu(T,S\setminus x)=0$ implies $\mu(T,S)+\mu(T\cup x, S)=0$
    \item [(ii)] $\mu(T,S\setminus x)>0$ implies $\mu(T,S)+\mu(T\cup x, S)>0$.
\end{itemize}
\end{lemma}

\begin{proof} For part (i), suppose $\mu(T^1,S\setminus x)=0$ with $T^1\subseteq S\setminus x$ and $x\in S$ and $T^1\ne \emptyset$. This implies there must exist a non-empty $T'\subseteq S\setminus x$ such that $\mu(T',S\setminus x)>0$. Plugging such $(T^1,T')$ into Relative Additivity (Axiom \ref{ax: RCG}):
\[
  \mu(T^1,S\setminus x) [\mu(T',S)+\mu(T'\cup x, S)]=\mu(T',S\setminus x) [\mu(T^1,S)+\mu(T^1\cup x, S)].
\]
The LHS of the equality above is equal to $0$ because $\mu(T^1,S\setminus x)=0$ by the selection of $T^1$. Hence, the RHS of the equality above is also $0$. As $\mu(T',S\setminus x)>0$ by the selection of $T'$, it follows $\mu(T^1,S)+\mu(T^1\cup x, S)=0$. 

For part (ii), suppose $\mu(T^2,S\setminus x)>0$ with $T^2\subseteq S\setminus x$ and $x\in S$ and $T^2\ne \emptyset$. We want to show $\mu(T^2,S)+\mu(T^2\cup x, S)>0$. Proof by contradiction. Suppose $\mu(T^2,S)+\mu(T^2\cup x, S)=0$. Take an arbitrary non-empty $T''\subseteq S\setminus x$. Plugging such $(T^2,T'')$ into Relative Additivity (Axiom \ref{ax: RCG}), we have
\[
  \mu(T^2,S\setminus x) [\mu(T'',S)+\mu(T''\cup x, S)]=\mu(T'',S\setminus x) [\mu(T^2,S)+\mu(T^2\cup x, S)].
\]
The RHS of the equality above is zero because $\mu(T^2,S)+\mu(T^2\cup x, S)=0$ by assumption. Hence, the LHS of the equality above must be $0$. As $\mu(T^2,S\setminus x)>0$, this implies that $\mu(T'',S)+\mu(T''\cup x, S)=0$. As $T''$ is chosen arbitrarily, this implies
\[
0=\sum_{T'':\emptyset \ne T''\subseteq S\setminus x}[\mu(T'',S)+\mu(T''\cup x,S)]=1-\mu(\{x\},S).
\]
The second equation comes from the fact that $\sum_{T: \emptyset \ne T\subseteq S} \mu(T,S)=1$ by the definition of $\mu$. Note that $1-\mu(\{x\},S)=0$ implies $\mu(\{x\},S)=1$. This contradicts Positivity-1 (Axiom \ref{ax: RCG-Positivity}). To elaborate, consider an arbitrary $y\in S\setminus x$. Such $y$ exists because $S\setminus x$ is non-empty. Then there does not exist a non-empty $T\subseteq S$ with $y\in T$ such that $\mu(T,S)>0$ because $\mu(\{x\},S)=1$ (contradiction). This completes our proof of Lemma \ref{lema: MA}.
\end{proof}
Back to the main proof of Proposition \ref{prop: RCG}. Suppose $\mu$ satisfies Positivity-1 and Relative Additivity. This implies that for all $x\in S\subseteq X$ and $T,T'\subseteq S\setminus x$ such that $T,T'\ne \emptyset$
\[
\mu(T,S\setminus x) [\mu(T',S)+\mu(T'\cup x, S)]=\mu(T',S\setminus x) [\mu(T,S)+\mu(T\cup x, S)].
\]
For each category $C\in \mathcal{X}$, define $m(C)=\mu(C,X)$. By Positivity-1 (Axiom \ref{ax: RCG-Positivity}), for every $x\in X$, there exists $T\ni x$ such that $\mu(T,X)>0$. It follows that $x$ belongs to at least one category $C$ with $m(C)>0$; the category $C$ can be chosen to be identical to $T$. Now, for all non-empty $T\subseteq S$, we show
\begin{equation}\label{eq: RCG*}
    \mu(T,S)=\sum_{C} \frac{m(C) \mathbbm{1}(T=S \cap C)}{\sum_{C: S\cap C\ne\emptyset} m(C)}=\frac{1}{\sum_{C: S\cap C\ne\emptyset} m(C)}\sum_{C:\,C=T\cup A \text{ with }A\subseteq X\setminus S} m(C)
\end{equation}
by induction based on the number of alternatives in $S$ by ``stepping down.''  

\noindent \underline{\textit{Step 1:}} When $S=X$, by definition $\displaystyle \mu(T,X)=m(T)=\sum_{C\in \mathcal{X}} m(C) \mathbbm{1}(T=X \cap C)$. Additionally, note that $\sum_{C: X\cap C\ne\emptyset} m(C)=\sum_{C: C\subseteq X, C\ne\emptyset} m(C)=\sum_{C: C\subseteq X, C\ne\emptyset}\mu(C,X)=1$. Hence, $\mu$ has the representation at $X$.

\noindent \underline{\textit{Step 2:}} Suppose equation (\ref{eq: RCG*}) holds for all $(T,S)$ such that $T\subseteq S$, $T\ne \emptyset$, and $|S|=k+1,k+2,\dots, |X|$. Consider a choice set $S$ with $x\in S$ and $|S|=k+1$. We show equation (\ref{eq: RCG*}) holds at $S\setminus x$. For notational simplicity, let $\beta_S=\sum_{C: S\cap C\ne\emptyset} m(C)$.

\underline{Case 2.1:} Take non-empty $T\subseteq S\setminus x$ such that $\mu(T,S\setminus x)=0$. It follows from Lemma \ref{lema: MA} that $\mu(T,S)+ \mu(T\cup x, S)=0$. Therefore, we can write $\mu(T,S\setminus x)=0=\mu(T,S)+\mu(T\cup x, S)$. This is equivalent to
\begin{eqnarray*}
    &&\underbrace{\frac{1}{\beta_S}\sum_{C:\,C=T\cup A \text{ with }A\subseteq X\setminus S} m(C)}_{\text{$\mu(T,S)$}}+\underbrace{\frac{1}{\beta_S}\sum_{C:\,C=(T\cup x)\cup A \text{ with } A\subseteq X\setminus S} m(C)}_{\text{$\mu(T\cup x, S)$}}\\
    &=& \frac{1}{\beta_S} \Bigg[\sum_{C:\,C=T\cup A',\, A'\subseteq ((X\setminus S)\cup x)  \text{ and } x\not \in A'} m(C) + \sum_{C:\,C=T\cup A',\, A'\subseteq ((X\setminus S)\cup x) \text{ and } x\in A'} m(C)\Bigg] \\
    &=& \frac{1}{\beta_S} \Bigg[\sum_{C:\,C=T\cup A',\,A'\subseteq X\setminus (S\setminus x)  \text{ and } x\not \in A'} m(C) + \sum_{C:\,C=T\cup A',\,A'\subseteq X\setminus (S\setminus x) \text{ and } x\in A'} m(C) \Bigg] \\
    &=&\frac{1}{\beta_S} \Bigg[\sum_{C:\,C=T\cup A',\,A'\subseteq X\setminus (S\setminus x)} m(C)\Bigg]=\frac{1}{\beta_{S\setminus x}} \Bigg[\sum_{C:\,C=T\cup A',\,A'\subseteq X\setminus (S\setminus x)} m(C)\Bigg].
\end{eqnarray*}
Note that we replace $\beta_S$ by $\beta_{S\setminus x}$ in the last equation. This does not impact the equality as the expression inside the big square bracket equals $0$ because $\mu(T,S)+ \mu(T\cup x, S)=\mu(T,S\setminus x)=0$ by the selection of $T$. Hence, equation (\ref{eq: RCG*}) holds for all non-empty $T\subseteq S\setminus x$ such that $\mu(T,S\setminus x)=0$. 

\underline{Case 2.2:} Fix a non-empty $T\subseteq S\setminus x$ such that $\mu(T,S\setminus x)>0$. Lemma \ref{lema: MA} implies that $\mu(T,S)+ \mu(T\cup x, S)>0$. Applying Relative Additivity with $(T,T')$, where non-empty $T'\subseteq S\setminus x$ can be chosen arbitrarily, we have
\begin{eqnarray*}
    \frac{\mu(T',S\setminus x)}{\mu(T,S\setminus x)}&=&\frac{\mu(T',S)+\mu(T'\cup x, S)}{\mu(T,S)+\mu(T\cup x, S)}=\frac{\sum_{C:\,C=T'\cup A \text{ with }A\subseteq X\setminus (S\setminus x)} m(C)}{\sum_{C:\,C=T\cup A \text{ with }A\subseteq X\setminus (S\setminus x)} m(C)},
\end{eqnarray*}
where the second equation follows a logic similar to that of case 2.1. Take a summation over all non-empty $T'\subseteq S\setminus x$
\begin{eqnarray*}
    \sum_{T':T'\subseteq S\setminus x, T'\ne \emptyset} \frac{\mu(T',S\setminus x)}{\mu(T,S\setminus x)} &=&\sum_{T':T'\subseteq S\setminus x, T'\ne \emptyset} \frac{\sum_{C:\,C=T'\cup A \text{ with }A\subseteq X\setminus (S\setminus x)} m(C)}{\sum_{C:\,C=T\cup A \text{ with }A\subseteq X\setminus (S\setminus x)} m(C)} \\
    \Leftrightarrow \frac{1}{\mu(T,S\setminus x)}&=&\frac{\beta_{S\setminus x}}{\sum_{C:\,C=T\cup A \text{ with }A\subseteq X\setminus (S\setminus x)} m(C)} \\
    \Leftrightarrow \mu(T,S\setminus x)&=&\frac{1}{\beta_{S\setminus x}}\sum_{C:\,C=T\cup A \text{ with }A\subseteq X\setminus (S\setminus x)} m(C),
\end{eqnarray*}
where the second equation uses $\sum_{T':T'\subseteq S\setminus x, T'\ne \emptyset}\mu(T',S\setminus x)=1$ and the definition of $\beta_{S\setminus x}$
\begin{eqnarray*}
    \beta_{S\setminus x}=\sum_{C: (S\setminus x)\cap C\ne\emptyset} m(C)= \sum_{T':\,T'\subseteq S\setminus x, T'\ne \emptyset} \Bigg[\sum_{C:\,C=T'\cup A \text{ with }A\subseteq X\setminus (S\setminus x)} m(C)\Bigg]
\end{eqnarray*}
Hence, equation (\ref{eq: RCG*}) holds for all non-empty $T\subseteq S\setminus x$ such that $\mu(T,S\setminus x)>0$. This completes our proof of Proposition \ref{prop: RCG}. $\blacksquare$

\bigskip

\noindent \textbf{Proof of Proposition \ref{prop: IC}:} The necessity can be easily verified. We show the sufficiency. As $\mu$ has full support and satisfies IIS, by Proposition \ref{prop: logit}, we can write $\mu(T,S)=\frac{\pi(T)}{\sum_{T':\emptyset \ne T'\subseteq S} \pi (T')}$ for all non-empty $T\subseteq S$, where $\pi(T)=\mu(T,X)$ for all non-empty $T\subseteq X$. Applying Relative Additivity (Axiom \ref{ax: RCG}) to this functional form of $\mu$ and using the full support of $\mu$, we have
\begin{equation}\label{eq: ICNoOutside2}
\frac{\pi(T)}{\pi (T')}=\frac{\pi(T)+\pi(T\cup x)}{\pi(T')+\pi(T'\cup x)} \Rightarrow \frac{\pi(T\cup x)}{\pi(T)}= \frac{\pi(T'\cup x)}{\pi(T')}
\end{equation}
for all non-empty $T,T'\subseteq X$ such that $x\not\in T,T'.$ Define a function $\gamma\colon X \rightarrow (0,1)$ such that $\gamma(x)=\frac{\pi(X)}{\pi(X)+\pi(X\setminus x)}$ for all $x\in X$. Equivalently, $\frac{\gamma(x)}{1-\gamma(x)}=\frac{\pi(X)}{\pi(X\setminus x)}$. We show $\frac{\pi(X)}{\pi (X\setminus T)}=\prod_{x\in T}\frac{\gamma(x)}{1-\gamma(x)}$ $\text{ for all non-empty } T\subseteq X$ such that $T\ne X$. Consider some $T=\{x_1,x_2,\dots,x_k\}$, where $x_i\ne x_j$ when $i\ne j$ such that $T\ne X$. Let $T_0=\emptyset$ and $T_i=\{x_1,\dots,x_i\}$ for $i=1,\dots,k$. By definition $T=T_k$. Equation (\ref{eq: ICNoOutside2}) implies that $\frac{\pi(X\setminus T_i )}{\pi (X\setminus T_{i+1})}=\frac{\pi(X )}{\pi (X\setminus x_{i+1})}=\frac{\gamma(x_{i+1})}{1-\gamma(x_{i+1})}$, where the second equation comes from the definition of $\gamma$. Note that $\frac{\pi(X)}{\pi (X\setminus T)}=\prod_{i=0}^{k-1} \frac{\pi(X\setminus T_i )}{\pi (X\setminus T_{i+1})}$ as $T_0=\emptyset$ and $T_k=T$. Hence, $\frac{\pi(X)}{\pi (X\setminus T)}=\prod_{i=0}^{k-1}\frac{\gamma(x_{i+1})}{1-\gamma(x_{i+1})}= \prod_{x\in T}\frac{\gamma(x)}{1-\gamma(x)}$ $\text{ for all non-empty } T\subseteq X$ such that $T\ne X$. 

Next, we express $\pi(X)$ in terms of $\gamma$. Note that 
$$\sum_{T:T\subseteq X, T\ne \emptyset}\pi(T)=\sum_{T:T\subseteq X, T\ne \emptyset} \mu(T,X)=1.$$
Therefore,
\[
\sum_{T:T\subset X, T\ne \emptyset}\frac{\pi (X\setminus T)}{\pi(X)}=\frac{1-\pi(X)}{\pi(X)} \Rightarrow \sum_{T:T\subset X,T\ne \emptyset } \prod_{x\in T}\frac{1-\gamma(x)}{\gamma(x)}=\frac{1-\pi(X)}{\pi(X)}.
\]
This yields $\pi (X)=\frac{\prod_{x\in X}\gamma(x)}{1-\prod_{x\in X}(1-\gamma(x))}$ using the binomial-style expansion. Since $\frac{\pi(X)}{\pi (X\setminus T)}=\prod_{x\in T}\frac{\gamma(x)}{1-\gamma(x)}$, it follows $\pi (T)=\frac{\prod_{x\in T}\gamma(x)\prod_{y\in X\setminus T}(1-\gamma(y))}{1-\prod_{x\in X}(1-\gamma(x))}$ for all non-empty $T\subseteq X$. Plugging this $\pi(T)$ into the logit representation of $\mu$ gives the IC representation. This completes our proof of Proposition \ref{prop: IC}. $\blacksquare$

\bigskip

\noindent \textbf{Proof of Remark \ref{rem: EBA-AR-RCG}:} First, suppose $\mu$ has an RCG representation. We define \textit{artificial aspects} based on $\mu$. At the grand set, enumerate all nonempty $T$ with positive probabilities of being chosen as $T_1,T_2,\dots,T_K$, where $T_i\ne T_j$ when $i\ne j$. Define $K$ artificial aspects by calling them aspect $1$, aspect 2,$\dots$, aspect $K$. Let the collection of options possessing artificial aspect $i$, $A_i$, be $A_i=T_i$. Let option $x$ possess attribute $i$ if and only if $x\in A_i$. By definition of the \text{artificial aspects} and the RCG representation of $\mu$, $\mu$ satisfies Positivity-2 (Axiom \ref{ax: EBA-posi}). Additionally, $\mu$ satisfies Relative Additivity because it has an RCG representation. It follows from Proposition \ref{prop: EBA} that $\mu$ has an EBA representation with the aspects we defined, which implies that it has an endogenous EBA representation. 

Now, suppose $\mu$ has an endogenous EBA representation. Let $\{i_1,\dots,i_M\}$ be the set of all (artificial) attributes. Let the set of options possessing attribute $i_t$ be $A_{i_t}$. Let each category $C$ be a set $A_{i_t}$. We show that $\mu$ has an RCG representation with these categories. First, we show that $\mu$ satisfies Positivity-1. By the definition of endogenous EBA, for an arbitrary option $x$, there exists an attribute $i_j$ such that $x$ possesses attribute $i_j$, i.e., $x\in A_{i_j}$. Hence, for each option $x$ the category $C=A_{i_j}$ includes $x$. This observation, together with the fact that $\mu$ satisfies Positivity-2 (because it has an endogenous EBA representation), implies that $\mu$ satisfies Positivity-1. Second, by Proposition \ref{prop: EBA}, $\mu$ also satisfies Relative Additivity since it has an endogenous EBA representation. Therefore $\mu$ satisfies both Positivity-1 and Relative Additivity, and it follows from Proposition \ref{prop: RCG} that $\mu$ has an RCG representation. This completes our proof of Remark \ref{rem: EBA-AR-RCG}. $\blacksquare$

\bigskip

\noindent \textbf{Proof of Proposition \ref{prop: RRM}:} First, we show the necessity of Axiom \ref{ax: RRM1}. The necessity of Axiom \ref{ax: RRM2} is similar so we omit a formal proof. 

\noindent \textit{\underline{Necessity of Axiom \ref{ax: RRM1}:}} Suppose $\mu$ has an RRM representation with $\{\mathcal{Q}(x)\}_x$ and $\{s_x\}_x$ being the constraint sets and salience parameters. Take $x\in S$ and $T\subseteq S\setminus x$ such that $T\ne \emptyset$. Let 
\begin{eqnarray*}
    C_1 &=& \{y: y\in S\setminus x \,\, \text{ such that } \, \, T=\mathcal{Q}(y)\cap (S\setminus x)\} \\
    C_2 &=& \{y:y\in S \,\, \text{ such that } \, \, T=\mathcal{Q}(y)\cap S\} \\
    C_3 &=& \{y:y\in S \,\, \text{ such that } \, \, T\cup x=\mathcal{Q}(y)\cap S\}.
\end{eqnarray*}
Clearly, $C_2$ and $C_3$ are disjoint. Using the definitions of $C_1,C_2,C_3$ and the RRM representation of $\mu$, we can write 
\begin{eqnarray*}
\mu(T,S\setminus x) &=& \bigg(\sum_{z\in S\setminus x}s_z\bigg)^{-1}\sum_y s_y\mathbbm{1}(y\in C_1) \\
\mu(T,S) &=& \bigg(\sum_{z\in S}s_z\bigg)^{-1}\sum_y s_y\mathbbm{1}(y\in C_2) \\
\mu(T\cup x, S) &=& \bigg(\sum_{z\in S}s_z\bigg)^{-1}\sum_y s_y\mathbbm{1}(y\in C_3).
\end{eqnarray*}
To show that Axiom \ref{ax: RRM1} is necessary, it is sufficient to show $C_1=C_2\cup C_3$ when $T \ne \mathcal{Q}(x)\cap (S\setminus x)$ and $T$ is non-empty. We do that in three steps. 

\textit{Step 1:} Take an arbitrary $y_1\in C_1$. It follows that $y_1\in S$. If $x\not\in \mathcal{Q}(y_1)$ then $T=\mathcal{Q}(y_1)\cap (S\setminus x)=\mathcal{Q}(y_1)\cap S$. Consequently, $y_1\in C_2$. If $x\in \mathcal{Q}(y_1)$ then $\mathcal{Q}(y_1)\cap S=T\cup x$. It follows that $y_1\in C_3$. Therefore, for each $y_1\in C_1$, either $y_1\in C_2$ or $y_1\in C_3$. It follows $C_1\subseteq C_2\cup C_3$. 

 \textit{Step 2:} Take an arbitrary $y_2\in C_2$. It follows that $y_2\in S$. First, note that $x\not\in \mathcal{Q}(y_2)$. This is because if $x\in \mathcal{Q}(y_2)$ then $x\in \mathcal{Q}(y_2)\cap S$ because $x\in S$. But then it follows that $x\in T=\mathcal{Q}(y_2)\cap S$, which contradicts $T\subseteq S\setminus x$. Second, $x\not\in \mathcal{Q}(y_2)$ implies that $T=\mathcal{Q}(y_2)\cap S=\mathcal{Q}(y_2)\cap (S\setminus x)$. Additionally, $x\not\in \mathcal{Q}(y_2)$ implies $y_2\ne x$ because $y_2\in \mathcal{Q}(y_2)$ by assumption. Hence, $y_2\in C_1$. 
    
\textit{Step 3:} Take an arbitrary $y_3\in C_3$. It follows that $y_3\in S$. By definition of $C_3$, the facts that $y_3\in C_3$ and $T\subseteq S\setminus x$ imply that $T=\mathcal{Q}(y_3)\cap (S\setminus x)$. Additionally, it must be the case that $y_3\ne x$. Suppose not and $y_3=x$. Then it implies that $T=\mathcal{Q}(x)\cap (S\setminus x)$, which contradicts the initial assumption that $T\ne \mathcal{Q}(x)\cap (S\setminus x)$. With $y_3\ne x$, $y_3\in S$, and $T=\mathcal{Q}(y_3)\cap (S\setminus x)$, it follows that $y_3\in C_1$. Combining the three steps, we have $C_1=C_2\cup C_3$.

\medskip
\noindent \textit{\underline{The sufficiency part:}} Suppose $\mu$ satisfies Axiom \ref{ax: RRM1} and Axioms \ref{ax: RRM-distinct}-\ref{ax: RRM2}. We show $\mu$ has an RRM representation. Define the constraint set $\mathcal{Q}(x)$ as the revealed constraint set $\mathcal{Q}^R(x)$. For each $x\in X$, define the salience parameter $s_x=\mu(\mathcal{Q}^R(x),X)$. Note that $s_x>0$ (because of Axiom \ref{ax: RRM-Positivity}) and $\sum_{x\in X}s_x=1$ because $\sum_{T:T\subseteq X, T\ne \emptyset} \mu(T,X)=1$. We first state and prove the following Lemma.

\begin{lemma}\label{lema:Axiom7} Axiom \ref{ax: RRM-Positivity} implies that for all $x,T,S$ with $x\in S$, $T\subseteq S\setminus x$, and $T\ne \emptyset$, if $\mu(T,S\setminus x)>0$ then $\mu(T,S)+\mu(T\cup x, S)>0$.
\end{lemma}

\begin{proof} Suppose $\mu(T,S\setminus x)>0$. Then Axiom \ref{ax: RRM-Positivity} implies that there exists $y\in S\setminus x$ such that $T= \mathcal{Q}^R(y)\cap (S\setminus x)$. If $x\in \mathcal{Q}^R(y)$ then $T\cup x= \mathcal{Q}^R(y)\cap S$. Therefore, Axiom \ref{ax: RRM-Positivity} implies that $\mu(T\cup x, S)>0$. Otherwise, if $x\not\in \mathcal{Q}^R(y)$ then $T= \mathcal{Q}^R(y)\cap S$. Axiom \ref{ax: RRM-Positivity} then implies that $\mu(T,S)>0$. In either case, $\mu(T,S)+\mu(T\cup x, S)>0$. This completes our proof of Lemma \ref{lema:Axiom7}.
\end{proof}

Back to the main proof of Proposition \ref{prop: RRM}. We show the sufficiency by induction based on the number of alternatives in $S$ by ``stepping down.'' Suppose $S=X$. First, consider a non-empty $T\subseteq X$ such that $\mu(T,X)=0$. Axiom \ref{ax: RRM-Positivity} implies that $T\ne \mathcal{Q}^R(x)\cap X$ for all $x\in X$. Hence, we can write 
\[
\mu(T,X)=0=\sum_{y\in X} \frac{s_y}{\sum_{z\in X} s_z}\mathbbm{1}(T=\mathcal{Q}^R(y)\cap X)
\]
Second, consider a non-empty $T\subseteq X$ such that $\mu(T,X)>0$. Axiom \ref{ax: RRM-Positivity} implies that there is at least one $y\in X$ such that $T=\mathcal{Q}^R(y)\cap X=\mathcal{Q}^R(y)$. Axiom \ref{ax: RRM-distinct} claims that such $y$ is unique. Hence, $\mu(T,X)=\mu(\mathcal{Q}^R(y),X)= s_{y} = \sum_{z\in X}s_z\mathbbm{1}(T=\mathcal{Q}^R(z)\cap X)$. The first two equations come from the definition of $y$ and parameters $\{s_{z}\}_{z\in X}$. The last equation comes from the uniqueness of $y$.

Now, suppose that $\mu$ has an RRM representation at all choice sets $S$ with $|S|=k+1,k+2,\dots, |X|$ with $k\ge 2$. Take an arbitrary $S$ such that $|S|=k+1$. Let $x$ be an arbitrary element in $S$. We show that $\mu$ also has an RRM representation at $S\setminus x$. First, take a non-empty $T\subseteq S\setminus x$ such that $\mu(T, S\setminus x)=0$. It follows from Axiom \ref{ax: RRM-Positivity} that $\nexists y\in S\setminus x$ such that $T=\mathcal{Q}^R(y)\cap (S\setminus x)$. Hence, we can write 
\[
\mu(T,S\setminus x)=0= \frac{\sum_{y\in S\setminus x} s_y \mathbbm{1}(T=\mathcal{Q}^R(y)\cap (S\setminus x))}{\sum_{z\in S\setminus x} s_z}.
\]
Second, take a non-empty $T\subseteq S\setminus x$ such that $\mu(T,S\setminus x)>0$. Consider the following cases.

\underline{\textit{Case 1:}} Suppose $\mathcal{Q}^R(x)\cap (S\setminus x)=\emptyset$. In this case, Axiom \ref{ax: RRM1} applies to all $T,T'\subseteq S\setminus x$ such that $T,T'\ne \emptyset$. As $\mu(T, S\setminus x)>0$, Lemma \ref{lema:Axiom7} implies that $\mu(T,S)+\mu(T\cup x, S)>0$. Hence, for all non-empty $T'\subseteq S\setminus x$
\begin{eqnarray*}
   \frac{\mu(T',S\setminus x)}{\mu(T,S\setminus x)}&=&\frac{\mu(T',S)+\mu(T'\cup x, S)}{\mu(T,S)+\mu(T\cup x, S)} \\
  &=& \frac{\sum_{y\in S} s_y \mathbbm{1}(T'=\mathcal{Q}^R(y)\cap S)+\sum_{y\in S} s_y \mathbbm{1}(T'\cup x=\mathcal{Q}^R(y)\cap S)}{\sum_{y\in S} s_y \mathbbm{1}(T=\mathcal{Q}^R(y)\cap S)+\sum_{y\in S} s_y \mathbbm{1}(T\cup x=\mathcal{Q}^R(y)\cap S) }
  \\
    &=&...(\text{following the same logic as in the necessary part of Axiom \ref{ax: RRM1})}\\
    &=& \frac{\sum_{y\in S\setminus x} s_y \mathbbm{1}(T'=\mathcal{Q}^R(y)\cap (S\setminus x))}{\sum_{y\in S\setminus x} s_y \mathbbm{1}(T=\mathcal{Q}^R(y)\cap (S\setminus x))}.
\end{eqnarray*}
Taking the summation over all non-empty $T'\subseteq S\setminus x$, we have
\begin{eqnarray*}
   \frac{\sum_{T': \emptyset \ne T'\subseteq S\setminus x}\mu(T',S\setminus x)}{\mu(T,S\setminus x)}&=& \frac{\sum_{T': \emptyset \ne T'\subseteq S\setminus x}\sum_{y\in S\setminus x} s_y \mathbbm{1}(T'=\mathcal{Q}^R(y)\cap (S\setminus x))}{\sum_{y\in S\setminus x} s_y \mathbbm{1}(T=\mathcal{Q}^R(y)\cap (S\setminus x))} \\
   \Leftrightarrow \frac{1}{\mu(T,S\setminus x)}&=& \frac{\sum_{y\in S\setminus x} s_y }{\sum_{y\in S\setminus x} s_y \mathbbm{1}(T=\mathcal{Q}^R(y)\cap (S\setminus x))} \\
   \Leftrightarrow \mu(T,S\setminus x)&=& \frac{\sum_{y\in S\setminus x} s_y \mathbbm{1}(T=\mathcal{Q}^R(y)\cap (S\setminus x))}{\sum_{y\in S\setminus x} s_y}.
\end{eqnarray*}
In the second equation above, the numerator of the ratio on the LHS is equal to $1$ by definition of the $\mu$ function. The numerator of the ratio on the RHS is equal to $\sum_{y\in S\setminus x}s_y$ because for each $y\in S\setminus x$, there exists a unique non-empty $T' \subseteq S\setminus x$ such that $T'=\mathcal{Q}^R(y)\cap (S\setminus x)$. The non-emptiness of such $T'$ comes from the fact that $y\in \mathcal{Q}^R(y)$ (by definition of the $\mathcal{Q}^R$ function) and $y\in S\setminus x$ (by the selection of $y$). The last equation implies that $\mu(T,S\setminus x)$ has an RRM representation at $S\setminus x$.

\bigskip
 \underline{\textit{Case 2:}} Suppose $\mathcal{Q}^R(x)\cap (S\setminus x)=T^*$ and $T^*\ne \emptyset$. In this case, Axiom \ref{ax: RRM1} is only applicable to all non-empty $T,T'\subseteq S\setminus x$ such that $T,T'\ne T^*$.

\textit{Case 2.1:} Suppose $\mu(T^*, S\setminus x)=1$. This implies $\mu(T,S\setminus x)=0$ for all non-empty $T\subseteq S\setminus x$ and $T\ne T^*$. Axiom \ref{ax: RRM-Positivity} implies that for all non-empty $T\subseteq S\setminus x$ such that $T\ne T^*$, $\nexists y\in S\setminus x$ such that $T=\mathcal{Q}^R(y)\cap (S\setminus x)$. Additionally, for all $z\in S\setminus x$, it must be the case that $\mathcal{Q}^R(z)\cap (S\setminus x)=T^*$. Otherwise, there exists another set $T^{**}=\mathcal{Q}^R(z)\cap (S\setminus x)$ that is non-empty (because $z\in S\setminus x$ and $z\in \mathcal{Q}^R(z))$ such that $\mu(T^{**},S\setminus x)>0$ and $T^{**}\ne T^*$, which is a contradiction. Hence, we can write 
\[
    \begin{cases}
    \mu(T^*,S\setminus x)&= \frac{\sum_{y\in S\setminus x} s_y \mathbbm{1}(T^*=\mathcal{Q}^R(y)\cap (S\setminus x))}{\sum_{z\in S\setminus x} s_z} \\
    \mu(T,S\setminus x)&=0= \frac{\sum_{y\in S\setminus x} s_y \mathbbm{1}(T=\mathcal{Q}^R(y)\cap (S\setminus x))}{\sum_{z\in S\setminus x} s_z}  \quad \text{for all $T\subseteq S\setminus x$ such that $T\ne \emptyset, T^*$.} \\
    \end{cases}
\]
The first equation comes from the fact that $\mathcal{Q}^R(z)\cap (S\setminus x)=T^*$ for all $z\in S\setminus x$. The second equation results from the fact that $\nexists y\in S\setminus x$ such that $\mathcal{Q}^R(y)\cap (S\setminus x)=T$ for all $T\subseteq S\setminus x$ such that $T\ne \emptyset$ and $T\ne T^*.$

\textit{Case 2.2:} Suppose $\mu(T^*, S\setminus x)<1$. Take non-empty $T\subseteq S\setminus x$ such that $\mu(T, S\setminus x)>0$ and $T\ne T^*$. Such $T$ must exist because the choice probabilities sum to 1. As $\mu(T, S\setminus x)>0$, it follows from Lemma \ref{lema:Axiom7} that $\mu(T,S)+\mu(T\cup x, S)>0$. Hence, for all $T'\subseteq S\setminus x$ such that $T'\ne \emptyset$ and $T'\ne T^*$
\begin{eqnarray*}
   \frac{\mu(T',S\setminus x)}{\mu(T,S\setminus x)}&=&\frac{\mu(T',S)+\mu(T'\cup x, S)}{\mu(T,S)+\mu(T\cup x, S)} \\
  &=& \frac{\sum_{y\in S} s_y \mathbbm{1}(T'=\mathcal{Q}^R(y)\cap S)+\sum_{y\in S} s_y \mathbbm{1}(T'\cup x=\mathcal{Q}^R(y)\cap S)}{\sum_{y\in S} s_y \mathbbm{1}(T=\mathcal{Q}^R(y)\cap S)+\sum_{y\in S} s_y \mathbbm{1}(T\cup x=\mathcal{Q}^R(y)\cap S) }
  \\
    &=&...(\text{following the same logic as in the necessary part of Axiom \ref{ax: RRM1})}\\
    &=& \frac{\sum_{y\in S\setminus x} s_y \mathbbm{1}(T'=\mathcal{Q}^R(y)\cap (S\setminus x))}{\sum_{y\in S\setminus x} s_y \mathbbm{1}(T=\mathcal{Q}^R(y)\cap (S\setminus x))}.
\end{eqnarray*}
Summing over all non-empty $T'\subseteq S\setminus x$ such that $T'\ne T^*$, we obtain
\[
   \frac{\sum_{T': \emptyset,T^* \ne T'\subseteq S\setminus x}\mu(T',S\setminus x)}{\mu(T,S\setminus x)}= \frac{\sum_{T': \emptyset,T^* \ne T'\subseteq S\setminus x}\sum_{y\in S\setminus x} s_y \mathbbm{1}(T'=\mathcal{Q}^R(y)\cap (S\setminus x))}{\sum_{y\in S\setminus x} s_y \mathbbm{1}(T=\mathcal{Q}^R(y)\cap (S\setminus x))},
\]
which can be simplified to
\begin{equation}\label{eq: RRM-1}
\frac{1-\mu(T^*,S\setminus x)}{\mu(T,S\setminus x)}= \frac{\sum_{y\in S\setminus x} s_y-\sum_{z\in S\setminus x} s_z \mathbbm{1}(T^*=\mathcal{Q}^R(z)\cap (S\setminus x))}{\sum_{y\in S\setminus x} s_y \mathbbm{1}(T=\mathcal{Q}^R(y)\cap (S\setminus x))}.
\end{equation}
Here, the numerator of the ratio on the LHS is equal to $1-\mu(T^*,S\setminus x)$ because we leave out $T^*$ when taking the summation. The same intuition applies to the numerator of the ratio on the RHS. Now, applying Axiom \ref{ax: RRM2} for $(T^*,T)$ with $\mu(T,S\setminus x)>0$ by selection of $T$ and $\mu(T,S)+\mu(T\cup x,S)>0$ by Lemma \ref{lema:Axiom7}, we have 
\[
\frac{\mu(T^*,S\setminus x)}{\mu(T,S\setminus x)}=\frac{\mu(T^*,S)+\mu(T^*\cup x, S)-\frac{\mu(\mathcal{Q}^R(x),X)}{\sum_{y\in S} \mu(\mathcal{Q}^R(y),X)}}{\mu(T,S)+\mu(T\cup x, S)}.
\]
By the definition of the parameters $\{s_x\}_{x\in X}$, the ratio $\frac{\mu(\mathcal{Q}^R(x),X)}{\sum_{y\in S} \mu(\mathcal{Q}^R(y),X)}$ is indeed $\frac{s_x}{\sum_{y\in S} s_y}$. Using the fact that $\mu$ has an RRM representation at choice set $S$, we have
\footnotesize
\begin{eqnarray*}
   &&\frac{\mu(T^*,S)+\mu(T^*\cup x, S)-\frac{\mu(\mathcal{Q}^R(x),X)}{\sum_{y\in S} \mu(\mathcal{Q}^R(y),X)}}{\mu(T,S)+\mu(T\cup x, S)} \\
   &=&\frac{\sum_{y\in S} s_y \mathbbm{1}(T^*=\mathcal{Q}^R(y)\cap S)+\sum_{y\in S} s_y \mathbbm{1}(T^*\cup x=\mathcal{Q}^R(y)\cap S)-s_x}{\sum_{y\in S} s_y \mathbbm{1}(T=\mathcal{Q}^R(y)\cap S)+\sum_{y\in S} s_y \mathbbm{1}(T\cup x=\mathcal{Q}^R(y)\cap S) } \\
   &=& \frac{\sum_{y\in S\setminus x} s_y \mathbbm{1}(T^*=\mathcal{Q}^R(y)\cap (S\setminus x))}{\sum_{y\in S\setminus x} s_y \mathbbm{1}(T=\mathcal{Q}^R(y)\cap (S\setminus x))}.
\end{eqnarray*}
\normalsize
Here, the second equation follows by the same reasoning as the necessity of Axiom \ref{ax: RRM1}, together with the fact that $T^*=\mathcal{Q}^R(x)\cap (S\setminus x)$. Therefore, 
\begin{equation}\label{eq: RRM-2}
\frac{\mu(T^*,S\setminus x)}{\mu(T,S\setminus x)}=\frac{\sum_{y\in (S\setminus x)} s_y \mathbbm{1}(T^*=\mathcal{Q}^R(y)\cap (S\setminus x))}{\sum_{y\in S\setminus x} s_y \mathbbm{1}(T=\mathcal{Q}^R(y)\cap (S\setminus x))}.
\end{equation}
Combining equations (\ref{eq: RRM-1}) and (\ref{eq: RRM-2}), we get 
\footnotesize
\[
\frac{1}{\mu(T,S\setminus x)}= \frac{\sum_{y\in S\setminus x} s_y}{\sum_{y\in S\setminus x} s_y \mathbbm{1}(T=\mathcal{Q}^R(y)\cap (S\setminus x))} \Leftrightarrow \mu(T,S\setminus x) = \sum_{y\in S\setminus x}\frac{s_y}{\sum_{z\in S\setminus x} s_z} \mathbbm{1}(T=\mathcal{Q}^R(y)\cap (S\setminus x)).
\]
\normalsize
Therefore, $\mu$ has an RRM representation at choice set $S\setminus x$ in all cases. On singleton menus, the representation follows immediately from normalization and $x\in \mathcal{Q}^R(x)$. By induction, $\mu$ has an RRM representation. This completes our proof of Proposition \ref{prop: RRM}. $\blacksquare$

\bigskip

\noindent \textbf{Proof of Proposition \ref{prop: NSC}:} The necessity is straightforward. We prove the sufficiency. Suppose the revealed nests are $N^R_1,N^R_2,\dots,N^R_q$. Define the nests as the revealed nests. If $q=1$  then $N^R_1=X$. It follows from Axiom \ref{ax: NSC-Positivity} that $\mu(S,S)=1$ for all non-empty $S\subseteq X$ and $\mu$ has an NSC representation for any weighting function $\sigma$. Suppose $q\ge 2$. Fix $x_1^*\in N^R_1$ and $x_2^*\in N^R_2$. Take an arbitrary non-empty $T_i\subseteq N^R_i$ for all $i=1,2,\dots,q$. Define the $\sigma$ function as follows. Assign $\sigma(\{x^*_1\})=1$. Let $\sigma(T_i)=\frac{\mu(T_i,T_i\cup x_1^*)}{\mu(\{x_1^*\},T_i\cup x_1^*)}$ for $i=2,3,\dots,q$ and $\sigma(T_1)=\frac{\mu(T_1,T_1\cup x_2^*)}{\mu(\{x_2^*\},T_1\cup x_2^*)}\frac{\mu(\{x_2^*\},\{x_1^*,x_2^*\})}{\mu(\{x_1^*\},\{x_1^*,x_2^*\})}$. Set $\sigma(\emptyset)=0$ and assign arbitrary strictly positive values to collections spanning different nests. These values do not enter the NSC formula. The $\sigma$ function is well-defined following Axiom \ref{ax: NSC-Positivity}.

Take arbitrary non-empty $T,T' \subseteq S$ such that $\mu(T,S)>0$ and $\mu(T',S)>0$. We want to show $\frac{\mu(T,S)}{\mu(T',S)}=\frac{\sigma(T)}{\sigma(T')}$. As $\mu(T,S)>0$ and $\mu(T',S)>0$, Axiom \ref{ax: NSC-Positivity} implies that $T=S\cap N_i^R$ and $T'=S\cap N^R_j$ for some values of $i$ and $j$. In general, $i$ and $j$ are not necessarily distinct. We consider the following cases. 

\textit{\underline{Case 1:}} Suppose $i,j\ne 1$. It follows $T\cap N_1^R=T'\cap N_1^R=\emptyset$. Let $T''=\{x_1^*\}, S'=T\cup x_1^*$, and $S''=T'\cup x_1^*$. Axiom \ref{ax: NSC-Positivity} implies that $\mu(T,S'),\mu(T'',S'),\mu(T'',S'')$ and $\mu(T',S'')$ are all positive. By applying Axiom \ref{ax: NSC-PIIS}, we have
\[
\frac{\mu(T,S)}{\mu(T',S)}=\frac{\mu(T,S')}{\mu(T'',S')}\frac{\mu(T'',S'')}{\mu(T',S'')}=\frac{\mu(T,T\cup x_1^*)}{\mu(\{x_1^*\},T\cup x_1^*)}\frac{\mu(\{x_1^*\},T'\cup x_1^*)}{\mu(T',T'\cup x_1^*)}=\frac{\sigma(T)}{\sigma(T')},
\]
where the second equation comes from the definitions of $T'',S',S''$ and the third equation uses the definition of $\sigma$. 

\textit{\underline{Case 2:}} Suppose $i=j=1$. It follows $T,T'\subseteq N_1^R$. Let $T''=\{x_2^*\}, S'=T\cup x_2^*$, and $S''=T'\cup x_2^*$. Again, Axiom \ref{ax: NSC-Positivity} implies that $\mu(T,S'),\mu(T'',S'),\mu(T'',S'')$ and $\mu(T',S'')$ are all positive. By applying Axiom \ref{ax: NSC-PIIS}, we have
\footnotesize
\begin{eqnarray*}
\frac{\mu(T,S)}{\mu(T',S)}=\frac{\mu(T,S')}{\mu(T'',S')}\frac{\mu(T'',S'')}{\mu(T',S'')}=\frac{\mu(T,T\cup x_2^*)}{\mu(\{x_2^*\},T\cup x_2^*)}\frac{\mu(\{x_2^*\},T'\cup x_2^*)}{\mu(T',T'\cup x_2^*)}=\frac{\sigma(T)\frac{\mu(\{x_1^*\},\{x_1^*,x_2^*\})}{\mu(\{x_2^*\},\{x_1^*,x_2^*\})}}{\sigma(T')\frac{\mu(\{x_1^*\},\{x_1^*,x_2^*\})}{\mu(\{x_2^*\},\{x_1^*,x_2^*\})}}=\frac{\sigma(T)}{\sigma(T')}.
\end{eqnarray*}
\normalsize
The first equation above uses Axiom \ref{ax: NSC-PIIS}. The second equation comes from the definitions of $T'',S'$, and $S''$. The third equation uses the definition of the $\sigma$ function. 

\textit{\underline{Case 3:}} Suppose $i=1\ne j$. Then $T\subseteq N_1^R$ and $T'\cap N_1^R=\emptyset$. Applying Axiom \ref{ax: NSC-PIIS}, with endpoints $T$ and $\{x_1^*\}$, we have
\begin{alignat*}{2}
\frac{\mu(T,T\cup x_2^*)}{\mu(\{x_2^*\},T\cup x_2^*)}\frac{\mu(\{x_2^*\},\{x_1^*,x_2^*\})}{\mu(\{x_1^*\},\{x_1^*,x_2^*\})}&=\frac{\mu(T,S)}{\mu(T',S)} \frac{\mu(T',T'\cup x_1^*)}{\mu(\{x_1^*\},T'\cup x_1^*)} \\
\Leftrightarrow \underbrace{\frac{\mu(T,T\cup x_2^*)}{\mu(\{x_2^*\},T\cup x_2^*)}\frac{\mu(\{x_2^*\},\{x_1^*,x_2^*\})}{\mu(\{x_1^*\},\{x_1^*,x_2^*\})}}_{\text{$\sigma(T)$}} \underbrace{\frac{\mu(\{x_1^*\},T'\cup x_1^*)}{\mu(T',T'\cup x_1^*)}}_{\text{$1/\sigma(T')$}}&=\frac{\mu(T,S)}{\mu(T',S)} \\
\Leftrightarrow \frac{\sigma(T)}{\sigma(T')}&=\frac{\mu(T,S)}{\mu(T',S)},
\end{alignat*}
where the second equation comes from the definition of the $\sigma$ function. 

Now, as $\frac{\mu(T,S)}{\mu(T',S)}=\frac{\sigma(T)}{\sigma(T')}$ for all $T,T',S$ such that $\mu(T,S)>0$ and $\mu(T',S)>0$, we have $\mu(T,S)=\frac{\mu(T,S)}{\sum_{T':\mu(T',S)>0}\mu(T',S)}=\frac{\sigma(T)}{\sum_{T':\mu(T',S)>0}\sigma(T')}=\sum_i \frac{\sigma(T)}{\sum_{j=1}^{q}\sigma(N_j\cap S)}\mathbbm{1}(T=N_i\cap S)$. Here, the last equation uses the fact that we define the nests as the revealed nests, and the revealed nests are pairwise disjoint. This completes our proof. $\blacksquare$

\bigskip

\noindent \textbf{Proof of Proposition \ref{prop: RRM-relationship}:} We sequentially show that part (i) is equivalent to part (ii) and part (iii).

\noindent \underline{\textit{Equivalence of Part (i) and Part (ii)}:} We show that $\mu$ has RRM and endogenous EBA representations if and only if it has an RRM representation with $\mathcal{Q}(x)=\{x\}$. The if part is straightforward. If $\mu$ has an RRM representation with $\mathcal{Q}(x)=\{x\}$ then $\mu$ is given by
\[
\begin{cases}
    \begin{array}{rll}
        \mu(\{x\}, S) & = \frac{s_x}{\sum_{y \in S} s_y} &\quad \text{for all } x \in S \subseteq X \\[2pt]
        \mu(T, S) & = 0 &\quad \text{for all } T \subseteq S \subseteq X \text{ and } |T| \ge 2.
    \end{array}
\end{cases}
\]
This $\mu$ has a singleton representation. It is routine to check that this $\mu$ satisfies Relative Additivity and Positivity-1. Hence, it follows from Proposition \ref{prop: RCG} and Remark \ref{rem: EBA-AR-RCG} that $\mu$ has an endogenous EBA representation. For the only if part, we proceed by contradiction. Suppose $\mu$ has RRM and endogenous EBA representations. Suppose there exists $x^*$ such that $\mathcal{Q}^R(x^*)\ne \{x^*\}$. Take an arbitrary $S$ such that $x^*\in S$ and $(S\setminus x^*)\cap \mathcal{Q}^R(x^*) \ne \emptyset$. Note that $\mu$ having an endogenous EBA representation implies that it satisfies Relative Additivity (Axiom \ref{ax: RCG}). That is, for all non-empty $T,T'\subseteq S\setminus x^*$ and $x^*\in S$, we have
\begin{equation}\label{eq: RRM-EBA:1}
 \mu(T,S\setminus x^*) [\mu(T',S)+\mu(T'\cup x^*, S)]=\mu(T',S\setminus x^*) [\mu(T,S)+\mu(T\cup x^*, S)].
 \quad 
\end{equation}
Meanwhile, $\mu$ having an RRM representation implies that it satisfies Axiom \ref{ax: RRM2}, which says that when $T=\mathcal{Q}^R(x^*) \cap (S\setminus x^*)$ and $T\ne \emptyset$ then for all non-empty $T'\subseteq S\setminus x^*$ such that $T'\ne T$
\footnotesize
\begin{equation}\label{eq: RRM-EBA:2}
\mu(T,S\setminus x^*) [\mu(T',S)+\mu(T'\cup x^*, S)]=\mu(T',S\setminus x^*) \Big[\mu(T,S)+\mu(T\cup x^*, S)-\frac{\mu(\mathcal{Q}^R(x^*),X)}{\sum_{y\in S} \mu(\mathcal{Q}^R(y),X)}\Big].
\end{equation}
\normalsize
Because of our assumption that $\mathcal{Q}^R(x^*)\ne \{x^*\}$ and by selection of $S$, such $T$ in equation (\ref{eq: RRM-EBA:2}) always exists. Note that $\frac{\mu(\mathcal{Q}^R(x^*),X)}{\sum_{y\in S} \mu(\mathcal{Q}^R(y),X)}>0$ because $\mu$ has an RRM representation (see Axiom \ref{ax: RRM-Positivity}). Hence, the two equations (\ref{eq: RRM-EBA:1}) and (\ref{eq: RRM-EBA:2}) can occur simultaneously if and only if $\mu(T',S\setminus x^*)=0$ for all non-empty $T'\subseteq S\setminus x^*$ such that $T'\ne T$. This implies that $\mu(T,S\setminus x^*)=\mu(\mathcal{Q}^R(x^*) \cap (S\setminus x^*),S\setminus x^*)=1$. Consequently, for all $y\in S\setminus x^*$ we have $\mathcal{Q}^R(y)\cap (S\setminus x^*)=\mathcal{Q}^R(x^*) \cap (S\setminus x^*)$. Otherwise, the decision maker chooses $\mathcal{Q}^R(y)\cap (S\setminus x^*)$ with a strictly positive probability under choice set $S\setminus x^*$, contradicting the fact that $\mu(\mathcal{Q}^R(x^*) \cap (S\setminus x^*),S\setminus x^*)=1$. 

Now take $S=X$. Using the argument above, for all $y\in X\setminus x^*$,
\[
\mathcal{Q}^R(y)\cap (X\setminus x^*)=
    \begin{cases}
      \mathcal{Q}^R(y)\setminus x^* & \text{if $x^*\in \mathcal{Q}^R(y)$}\\
      \mathcal{Q}^R(y) & \text{if $x^*\not\in \mathcal{Q}^R(y)$}.
    \end{cases}       
\]
Note that $\mathcal{Q}^R(x^*)\cap (X\setminus x^*)=\mathcal{Q}^R(x^*)\setminus x^*$. Therefore, using $\mathcal{Q}^R(y)\cap (X\setminus x^*)=\mathcal{Q}^R(x^*) \cap (X\setminus x^*)$ for all $y\in X\setminus x^*$ (note that $S=X$ here) we have
\begin{equation}\label{eq: RRM-EBA:3}
\mathcal{Q}^R(x^*)\setminus x^*=
    \begin{cases}
      \mathcal{Q}^R(y)\setminus x^* & \text{if $x^*\in \mathcal{Q}^R(y)$}\\
      \mathcal{Q}^R(y) & \text{if $x^*\not\in \mathcal{Q}^R(y)$}.
    \end{cases}       
\end{equation}
When $|X|\ge 3$, there are at least two alternatives $y,z\in X\setminus x^*$ such that $y\ne z$. If
$x^*\in \mathcal{Q}^R(y)$ then (\ref{eq: RRM-EBA:3}) implies that $\mathcal{Q}^R(x^*)\setminus x^*= \mathcal{Q}^R(y)\setminus x^*$, which then implies  $\mathcal{Q}^R(y)= \mathcal{Q}^R(x^*)$. This contradicts the assumption that constraint sets associated with two different options are distinct. Now, suppose $x^*\not\in \mathcal{Q}^R(y)$ and $x^*\not\in \mathcal{Q}^R(z)$. Then it follows from (\ref{eq: RRM-EBA:3}) that $\mathcal{Q}^R(x^*)\setminus x^*=\mathcal{Q}^R(y)$ and $\mathcal{Q}^R(x^*)\setminus x^*=\mathcal{Q}^R(z)$. Consequently, $\mathcal{Q}^R(y)=\mathcal{Q}^R(z)$, which is also a contradiction. Hence, the initial assumption is wrong and $\mathcal{Q}^R(x)=\{x\}$ for all $x\in X$.  

Now, the equivalence of parts (i) and (ii) immediately follows from the fact that $\mu$ has a singleton representation if and only if it has an RRM representation with $\mathcal{Q}(x)=\{x\}$.

\smallskip
\noindent \underline{\textit{Equivalence of Part (i) and Part (iii)}:} First, suppose $\mu$ has a singleton representation. To obtain an NSC representation of $\mu$, let each nest contain only one alternative, so there are $|X|$ nests in total. To obtain an RRM representation of $\mu$, let the constraint associated with option $x$ be $\mathcal{Q}(x)=\{x\}$. Define the weighting function $\sigma$ in NSC and salience parameter $s_x$ in RRM as $\sigma(\{x\})=s_x=\mu(\{x\},X)$ for all $x\in X$. Set $\sigma(\emptyset)=0$ and assign arbitrary strictly positive values to all nonsingleton collections. It is routine to check that with the collection of nests $\{\{x\},\{y\},\{z\},\dots\}$ and weighting function $\sigma$, $\mu$ has an NSC representation. Additionally, with the constraint sets $\{\mathcal{Q}(x)\}_x$ and salience parameters $\{s_x\}_x$ defined as above, $\mu$ has an RRM representation.

Now, suppose $\mu$ has RRM and NSC representations. Consider choice probabilities at the grand set. Note that Axiom \ref{ax: RRM-Positivity} in RRM states that $\mu(T,X)>0$ if and only if $T=\mathcal{Q}^R(x)$ for some value of $x$, where $\mathcal{Q}^R()$ is the revealed constraint function. Meanwhile, Axiom \ref{ax: NSC-Positivity} in NSC states that $\mu(T,X)>0$ if and only if $T=N_i^R$ for some value of $i$, where $N_i^R$ is a \textit{revealed nest}. It follows from these two axioms that each revealed constraint set $\mathcal{Q}^R(x)$ identically maps to one revealed nest $N_i^R$. Note that for each $x\in X$ there is one $\mathcal{Q}^R(x)$. Also, the constraint sets associated with two different options are assumed to be distinct. Furthermore, the revealed nests are pairwise disjoint and must form a partition of the grand set (Axiom \ref{ax: NSC-Partition} in NSC). This implies that there are $|X|$ distinct revealed nests $N^R_1,N^R_2,\dots,N^R_{|X|}$ with $N_i^R=\{x\}$ for some $x\in X$. Using this result, it follows from Axiom \ref{ax: NSC-Positivity} (or Axiom \ref{ax: RRM-Positivity}) that at any choice set $S$, $\mu(\{x\},S)>0$ for all $x\in S$ and $\mu(T,S)=0$ when $|T|\ge 2$. Additionally, it follows from the Path-Independence of Irrelevant Sets (Axiom \ref{ax: NSC-PIIS} in NSC) that $\frac{\mu(\{x\},S)}{\mu(\{y\},S)}=\frac{\mu(\{x\},S')}{\mu(\{y\},S')}$ for all $S,S'\supseteq \xy$. Hence, $\mu$ satisfies both conditions in Definition \ref{defn: Singleton-mu}, which implies that it has a singleton representation. This completes our proof of Proposition \ref{prop: RRM-relationship}. $\blacksquare$


\smallskip
\noindent \textbf{Proof of Proposition \ref{prop: NSC-relationship}:} We sequentially show that part (i) is equivalent to part (ii) and part (iii).

\noindent \underline{\textit{Equivalence of Part (i) and Part (ii):}} First, suppose $\mu$ has a nest-invariant representation. Clearly $\mu$ has an NSC representation. To show that $\mu$ has an endogenous EBA representation, it is sufficient to show that $\mu$ satisfies Positivity-1 (Axiom \ref{ax: RCG-Positivity}) and Relative Additivity (Axiom \ref{ax: RCG}). It is straightforward to see that $\mu$ satisfies Positivity-1. For Relative Additivity, we want to show that for all $x,T,T',S$ such that $x\in S$ and $T,T' \subseteq S\setminus x$ and $T,T'\ne \emptyset$
\begin{equation}\label{eq: NSC-relationship}
    \mu(T,S\setminus x) [\mu(T',S)+\mu(T'\cup x, S)] = \mu(T',S\setminus x) [\mu(T,S)+\mu(T\cup x, S)].
\end{equation}
If $\mu(T,S\setminus x)=0$, then the NSC representation of $\mu$ implies that there does not exist a nest $N_i$ such that $T=N_i\cap (S\setminus x)$. This observation, together with the fact that $x\in S\setminus T$, implies that there does not exist a nest $N_i$ such that $T=N_i\cap S$ or $T\cup x=N_i\cap S$. It follows $\mu(T,S)=\mu(T\cup x,S)=0$ so $0=\mu(T,S\setminus x)=\mu(T,S)+\mu(T\cup x,S)$. In this case, (\ref{eq: NSC-relationship}) holds for any non-empty $T'\subseteq S\setminus x$. 

Now, suppose $\mu(T,S\setminus x)>0$ and $\mu(T',S\setminus x)>0$. The NSC representation of $\mu$ implies that there exist nests $N_i$ and $N_j$ (not necessarily distinct) such that $T=N_i\cap (S\setminus x)$ and $T'=N_j\cap (S\setminus x)$. Consider the following cases.

\medskip
\textit{Case 1:} Suppose $x\in N_i$ and $x\in N_j$. It follows that $N_i$ and $N_j$ are identical. In this case, $T$ and $T'$ are identical and (\ref{eq: NSC-relationship}) holds trivially.

\medskip
\textit{Case 2:} Suppose $x\in N_i$ and $x\not \in N_j$ (the case $x\not \in N_i$ and $x \in N_j$ is similar). It follows that $T\cup x=N_i\cap S$ and $T'=N_j\cap S$. Consequently, $\mu(T,S)=\mu(T'\cup x,S)=0$, $\mu(T\cup x,S)>0$ and $\mu(T',S)>0$. Equation (\ref{eq: NSC-relationship}) is then equivalent to
\[
\frac{\mu(T,S\setminus x)}{\mu(T',S\setminus x)} =  \frac{\mu(T\cup x,S)}{\mu(T',S)} \Leftrightarrow \frac{\sigma(T)}{\sigma(T')} =  \frac{\sigma(T\cup x)}{\sigma(T')}, 
\]
which is true because $\sigma(T)=\sigma(T\cup x)$ as both $T$ and $T\cup x$ belong to the nest $N_i$.

\medskip
\textit{Case 3:} Suppose $x\not\in N_i$ and $x\not\in N_j$. It follows that $T=N_i\cap S$ and $T'=N_j\cap S$. Consequently, $\mu(T\cup x,S)=\mu(T'\cup x,S)=0$, $\mu(T,S)>0$ and $\mu(T',S)>0$. Equation (\ref{eq: NSC-relationship}) is then equivalent to
\[
\frac{\mu(T,S\setminus x)}{\mu(T',S\setminus x)} =  \frac{\mu(T,S)}{\mu(T',S)} \Leftrightarrow \frac{\sigma(T)}{\sigma(T')} =  \frac{\sigma(T)}{\sigma(T')}, 
\]
which is true.

For the only if part, suppose $\mu$ has NSC and endogenous EBA representations. Suppose the nests in the NSC representation are $N_1,N_2,\dots, N_q$. If $q=1$ then there is only one nest being identical to the grand set: $N_1=X$. It follows that $\mu(S,S)=1$ for all non-empty $S\subseteq X$ regardless of the weighting function $\sigma$. As the weighting function $\sigma$ can be chosen arbitrarily, we can select $\sigma$ such that $\sigma(T)=\sigma(X)$ for all non-empty $T\subseteq X$. It follows that $\mu$ has a nest-invariant representation.

Now, suppose $q\ge 2$ so there are at least two distinct nests. We show that $\sigma(T)=\sigma(N_i)$ for all non-empty $T\subseteq N_i$. Note that this trivially holds when nest $N_i$ consists of a single option. Suppose there are at least two elements in $N_i$. Take an arbitrary non-empty subset $T$ of $N_i$ such that $T\ne N_i$. As $T$ is a strict subset of $N_i$, there exists $x \in N_i$ such that $T\subseteq N_i\setminus x$. Take an arbitrary non-empty $T'\subseteq N_j$ for $j\ne i$. We can always choose such $T'$ as there are at least two nests. By selection, $T$ and $T'$ are in different nests. Consider choice set $S=T\cup \{x\} \cup T'$. Note that as $\mu$ has an NSC representation, it follows from Axiom \ref{ax: NSC-Positivity} that all four terms $\mu(T,S\setminus x),\mu(T',S\setminus x),\mu(T\cup x,S),$ and $\mu(T',S)$ are strictly positive and $\mu(T'\cup x,S)=\mu(T,S)=0$. As $\mu$ has an endogenous EBA representation, it satisfies Relative Additivity (Axiom \ref{ax: RCG}). This implies
\begin{eqnarray*}
    \mu(T,S\setminus x)[\mu(T',S)+\mu(T'\cup x,S)]&=&\mu(T',S\setminus x)[\mu(T,S)+\mu(T\cup x,S)] \\
\Rightarrow  \mu(T,S\setminus x)\mu(T',S)&=&\mu(T',S\setminus x)\mu(T\cup x,S) \\
\Leftrightarrow  \frac{\mu(T,S\setminus x)}{\mu(T',S\setminus x)}&=&\frac{\mu(T\cup x,S)}{\mu(T',S)} \\
\Leftrightarrow  \sigma(T)/\sigma(T')&=&\sigma(T\cup x)/\sigma(T'),
\end{eqnarray*}
where the second equation uses $\mu(T'\cup x,S)=\mu(T,S)=0$, the third equation uses $\mu(T',S\setminus x)>0$, $\mu(T',S)>0$, and the last equation uses the NSC representation of $\mu$. Note that the last equation implies that $\sigma(T)=\sigma(T\cup x)$. As we choose $T$ and $x$ arbitrarily, by applying this process iteratively, we have $\sigma(T)=\sigma(N_i)$ for all non-empty $T\subseteq N_i$.

\smallskip
\noindent \underline{\textit{Equivalence of Part (i) and Part (iii):}} First, suppose $\mu$ has a nest-invariant representation. This implies that $\sigma(T)=\sigma(T')$ if $T$ and $T'$ belong to the same nest and $T,T'\ne \emptyset$. Take arbitrary $x,T,S$ such that $x\in S\setminus T$ and $T\subseteq S\setminus x$ and $T\ne \emptyset$. We want to show that $\mu$ satisfies the probabilistic attention filter condition; that is, $\mu(T,S)=\mu(T,S\setminus x)$ when $\mu(T,S)>0$ and $\mu(\{x\},S)=0$. Note that $\mu(T,S)>0$ implies that there exists a nest $N_i$ with $T=N_i\cap S$. Obviously, $x\not\in N_i$ because otherwise $x\in N_i\cap S$ and this implies $x\in T$ (contradiction). It follows that $N_i\cap (S\setminus x)=T$. Using the NSC representation of $\mu$, we have 
\[
\mu(T,S)=\frac{\sigma(N_i\cap S)}{\sum_j \sigma(N_j\cap S)} \quad \text{and} \quad \mu(T,S\setminus x)=\frac{\sigma(N_i\cap (S\setminus x))}{\sum_j \sigma(N_j\cap (S\setminus x))}.
\]
As $N_i\cap S=T=N_i\cap (S\setminus x)$, we have $\sigma(N_i\cap S)=\sigma(N_i\cap (S\setminus x))$. Hence, to show that $\mu(T,S)=\mu(T,S\setminus x)$, it is sufficient to prove $\sum_j \sigma(N_j\cap S)=\sum_j \sigma(N_j\cap (S\setminus x))$. We show $\sigma(N_j\cap S)=\sigma(N_j\cap (S\setminus x))$ for each nest $N_j$. Consider the following cases. 

\medskip
\textit{Case 1:} Suppose $N_j\cap S=N_j\cap (S\setminus x)=\emptyset$. Then $\sigma(N_j\cap S)=\sigma(N_j\cap (S\setminus x))$ trivially holds. 

\medskip
\textit{Case 2:} Suppose $N_j\cap S\ne \emptyset$ and $ N_j\cap (S\setminus x)\ne \emptyset$. Then both $N_j\cap S$ and $ N_j\cap (S\setminus x)$ are non-empty subsets of $N_j$. Using the fact that $\sigma$ is invariant across non-empty subsets of the same nest, we have $\sigma(N_j\cap S)=\sigma(N_j\cap (S\setminus x))$. 

\medskip
\textit{Case 3:} Suppose $N_j\cap S\ne \emptyset$ but $ N_j\cap (S\setminus x)= \emptyset$. Note that $N_j\cap S\ne \emptyset$ but $N_j\cap (S\setminus x)= \emptyset$ implies $N_j=E \cup x$ for some $E\subseteq X\setminus S$. Hence, it follows that $N_j\cap S=\{x\}$ and $\mu(\{x\},S)>0$. This contradicts the initial assumption that $\mu(\{x\},S)=0$. Therefore, the case in which $N_j\cap S\ne \emptyset$ but $ N_j\cap (S\setminus x)= \emptyset$ cannot happen.

\medskip
\textit{Case 4:} Suppose $N_j\cap S= \emptyset$ but $ N_j\cap (S\setminus x)\ne \emptyset$. This case cannot happen because $N_j\cap (S\setminus x)$ is a subset of $N_j\cap S$.

\medskip
Now, suppose $\mu$ has an NSC representation and satisfies the probabilistic attention filter condition. We want to show that $\mu$ has a nest-invariant representation. If there is just one nest $N_1=X$ then the NSC representation of $\mu$ implies that $\mu(T,S)=0$ when $T\ne S$ and $\mu(S,S)=1$ for all non-empty $S\subseteq X$. Choosing $\sigma(T)=1$ for every nonempty $T\subseteq X$, with $\sigma(\emptyset)=0$, therefore gives a nest-invariant representation of the same correspondence. Now, suppose that there exist at least two distinct nests, $N_i$ and $N_j$. We show that $\sigma(T)$ is the same for all non-empty $T\subseteq N_i$. This trivially holds if $N_i$ contains a single option. Suppose $N_i$ has at least two elements. Consider $x\in N_i$ and non-empty $T,T'$ such that $T\subseteq N_i\setminus x$ and $T'\subseteq N_j$. Let $S=T\cup \{x\}\cup T'$. We have $\mu(T',S)>0$ and $\mu(T',S\setminus x)>0$ because $T'=N_j\cap S=N_j\cap (S\setminus x)$. Additionally, $\mu(\{x\},S)=0$ because $x\in N_i$ but $N_i\cap S=T\cup x\ne \{x\}$. Then the probabilistic attention filter condition is applicable and implies that $\mu(T',S)=\mu(T',S\setminus x)$. Using the NSC representation of $\mu$ and the definition of $(T,T',S)$, 
\[
\mu(T',S)=\mu(T',S\setminus x) \quad \text{ implies } \quad \frac{\sigma(T')}{\sigma(T')+\sigma(N_i\cap S)}= \frac{\sigma(T')}{\sigma(T')+\sigma(N_i\cap (S\setminus x))}.
\]
Note that $\sigma(T')>0$ as $T'\ne \emptyset$. Consequently, the equation above implies $\sigma(N_i\cap S)=\sigma(N_i\cap (S\setminus x))$. Equivalently, $\sigma(T\cup x)=\sigma(T)$. As $x$ and $T$ are chosen arbitrarily, applying a similar logic iteratively implies that $\sigma(T)$ must be the same for all non-empty $T\subseteq N_i$. This completes our proof of Proposition \ref{prop: NSC-relationship}. $\blacksquare$

\bigskip

\noindent \textbf{Proof of Proposition \ref{prop: RCG*}:} The necessity part is straightforward. We prove the sufficiency. For each category $C\subseteq X$, define $m(C)=\mu(C,X)$. For all $T\subseteq S$, we show 
\begin{equation}\label{eq: RCG}
    \mu(T,S)=\sum_{C:\,C\subseteq X} m(C)\mathbbm{1}(T=S\cap C)=\sum_{C:\,C=T\cup A \text{ with }A\subseteq X\setminus S} m(C)
\end{equation}
by induction based on the number of alternatives in $S$ by ``stepping down''.

\noindent \underline{\textit{Step 1:}} When $S=X$, by definition $ \mu(T,X)=m(T)= \sum_{C\subseteq X} m(C)\mathbbm{1}(T=X\cap C)$ because $X \cap C=C$ for all categories $C$.

\noindent \underline{\textit{Step 2:}} Suppose equation (\ref{eq: RCG}) holds for all $(T,S)$ with $T\subseteq S$ and $|S|=k+1,k+2,\dots, |X|$, where $k\ge 1$. Consider some $S$ with $|S|=k+1$. Take $x\in S$. By Axiom \ref{ax: RCG*}, for all $T\subseteq S\setminus x$, we have
\begin{eqnarray*}
   \mu(T,S\setminus x)&=&\mu(T,S)+\mu(T\cup x, S)    \\
    &=&  \underbrace{\sum_{C:\,C=T\cup A \text{ with }A\subseteq X\setminus S} m(C)}_{\text{$\mu(T,S)$}}+\underbrace{\sum_{C:\,C=(T\cup x)\cup A,\,A\subseteq X\setminus S} m(C)}_{\text{$\mu(T\cup x, S)$}}\\
    &=& \sum_{C:\,C=T\cup A',\,A'\subseteq ((X\setminus S)\cup x),\,x\not \in A'} m(C) + \sum_{C:\,C=T\cup A',\,A'\subseteq ((X\setminus S)\cup x),\,x\in A'} m(C) \\
    &=& \sum_{C:\,C=T\cup A',\,A'\subseteq X\setminus (S\setminus x),\,x\not \in A'} m(C) + \sum_{C:\,C=T\cup A',\,A'\subseteq X\setminus (S\setminus x),\, x\in A'} m(C) \\
    &=&\sum_{C:\,C=T\cup A',\,A'\subseteq X\setminus (S\setminus x)} m(C).
\end{eqnarray*}
Hence, equation (\ref{eq: RCG}) holds at choice set $S\setminus x$. By induction, $\mu$ has a representation at any non-empty choice set. This completes our proof of Proposition \ref{prop: RCG*}. $\blacksquare$

\bigskip

\noindent \textbf{Proof of Proposition \ref{prop: IC*}:} The necessity can be easily verified. We show the sufficiency. As $\mu$ has full support and satisfies IIS$^o$, by a logic similar to that in the proof of Proposition \ref{prop: logit}, we can write $\mu(T,S)=\frac{\pi(T)}{\sum_{T'\subseteq S} \pi (T')}\,\, \text{for all } T\subseteq S\subseteq X$, where $\pi(T)=\mu(T,X)>0$ because $\mu$ has full support. Applying Additivity (Axiom \ref{ax: RCG*}) to this functional form of $\mu$, for all $T,T'\subseteq X$ such that $x\not\in T,T'$, we have
\begin{eqnarray*}
\frac{\mu(T,X\setminus x)}{\mu(T',X\setminus x)}=\frac{\mu(T,X)+\mu(T\cup x ,X)}{\mu(T',X)+\mu(T'\cup x ,X)} \Leftrightarrow \frac{\pi(T)}{\pi (T')}&=&\frac{\pi(T)+\pi(T\cup x)}{\pi(T')+\pi(T'\cup x)}  \\
\Rightarrow \frac{\pi(T\cup x)}{\pi(T)}&=& \frac{\pi(T'\cup x)}{\pi(T')}.
\end{eqnarray*}
The last equation is equation (\ref{eq: ICNoOutside2}) in the proof of Proposition \ref{prop: IC}. From here one follows the same logic used in the proof of Proposition \ref{prop: IC} to obtain the representation for the IC$^o$ model. This completes our proof.  $\blacksquare$

\medskip

\noindent \textbf{Proof of Corollary \ref{coro: nested logit}:} Suppose $\mu$ has a nested logit representation. If each nest in the nested logit representation of $\mu$ consists of a single item, then $\mu$ has a singleton representation following the proof of Proposition \ref{prop: RRM-relationship}. Suppose there exists a nest $N_i$ containing at least two elements. Take a proper subset $T$ of $N_i$ with $T\ne \emptyset$. As the utility of an option is strictly positive, it follows that $\sigma(T)\ne \sigma(N_i)$. Hence, $\mu$ does not have a nest-invariant representation. The remaining results in the Corollary directly follow from Propositions \ref{prop: RRM-relationship} and \ref{prop: NSC-relationship}. $\blacksquare$

\end{appendices}
\end{document}